\documentclass{fundam}
\usepackage{amsfonts}
\usepackage{amssymb}
\usepackage{amsmath}
\usepackage{graphicx}
\usepackage{hyperref}
\newcommand*{\overarrow}[1]{\overset{#1}\Longrightarrow}
\newcommand*{\overunderarrow}[2]{\underset{#2}{\overset{#1}\Longrightarrow}}
\newcommand*{\pref}{\qopname\relax m{Pref}}

\newcommand*{\buchi}{B\"{u}chi }
\renewcommand*{\inf}{\qopname\relax m{inf}}
\newcommand*{\ran}{\qopname\relax m{ran}}
\newcommand*{\wHyp}[1]{\omega\text{-}\qopname\relax m{#1}}
\newcommand*{\HypwHyp}[2]{\qopname\relax m{#1}\text{-}\omega\text{-}\qopname\relax m{#2}}
\newcommand*{\mwHyp}[1]{$m$\text{-}\omega\text{-}\qopname\relax m{#1}}
\newcommand*{\wHypV}[1]{\omega\text{-}\qopname\relax m{#1}\text{-}\qopname\relax m{V}}
\begin{document}
\setcounter{page}{119}
\issue{111~(2011)}

\title{On the Generative Power of $\omega$-Grammars and $\omega$-Automata}

\address{College of Computer Science and Technology, Nanjing University of Aeronautics and Astronautics, 29 Yudao Street, 210016 Nanjing, Jiangsu, China}

\author{Zhe Chen\\
(1) College of Computer Science and Technology\\
~~~~~~Nanjing University of Aeronautics and Astronautics\\
~~~~~~29 Yudao Street, 210016 Nanjing, Jiangsu, China\\
~~~~~~zhechen{@}nuaa.edu.cn\\
(2) LAAS-CNRS, INSA\\
~~~~~~Universit\'{e} de Toulouse\\
~~~~~~135 Avenue de Rangueil, 31077 Toulouse, France\\
~~~~~~zchen{@}insa-toulouse.fr }
\maketitle

\runninghead{Z. Chen}{On the Generative Power of $\omega$-Grammars and $\omega$-Automata}

\begin{abstract}
An $\omega$-grammar is a formal grammar used to generate $\omega$-words (i.e. infinite length words), while an $\omega$-automaton is an automaton used to recognize $\omega$-words. This paper gives clean and uniform definitions for $\omega$-grammars and $\omega$-automata, provides a systematic study of the generative power of $\omega$-grammars with respect to $\omega$-automata, and presents a complete set of results for various types of $\omega$-grammars and acceptance modes. We use the tuple $(\sigma,\rho,\pi)$ to denote various acceptance modes, where $\sigma$ denotes that some designated elements should appear at least once or infinitely often, $\rho$ denotes some binary relation between two sets, and $\pi$ denotes normal or leftmost derivations. Technically, we propose $(\sigma,\rho,\pi)$-accepting $\omega$-grammars, and systematically study their relative generative power with respect to $(\sigma,\rho)$-accepting $\omega$-automata. We show how to construct some special forms of $\omega$-grammars, such as $\epsilon$-production-free $\omega$-grammars. We study the equivalence or inclusion relations between $\omega$-grammars and $\omega$-automata by establishing the translation techniques. In particular, we show that, for some acceptance modes, the generative power of $\wHyp{CFG}$ is strictly weaker than $\wHyp{PDA}$, and the generative power of $\wHyp{CSG}$ is equal to $\wHyp{TM}$ (rather than linear-bounded $\omega$-automata-like devices). Furthermore, we raise some remaining open problems for two of the acceptance modes.
\end{abstract}


\section{Introduction}
\label{Sec:Intro}
An $\omega$-language is a set of $\omega$-words (i.e. infinite length words) over some alphabet $\Sigma$. An $\omega$-grammar is a formal grammar used to generate $\omega$-words, while an $\omega$-automaton is an automaton used to recognize $\omega$-words. The theory of $\omega$-languages has been studied in the literature in various formalisms. Most of the works focus on two aspects.

The first one is the relationship between $\omega$-automata and the theory of second order logic, and related decision problems. \buchi \cite{Buc60} started the study of obtaining decision procedures for some theory of restricted second order logic by using finite state $\omega$-automata. After that, a number of papers \cite{Buc65,BL69,ER66,Rab69} continued the discussion by examining the relationship between various formalisms of $\omega$-automata and the theory of second order logic. Thomas summarized related work in \cite{Tho91,Tho97}.

The second aspect concerns the generative power of $\omega$-automata and $\omega$-grammars, and the closure property of $\omega$-languages. McNaughton \cite{McN66} investigated finite state $\omega$-automata with various acceptance modes, and proved the equivalences between these variants, leading to the characterization of regular $\omega$-languages. Landweber \cite{Lan69} classified the families of $\omega$-languages accepted by deterministic finite state $\omega$-automata in the Borel hierarchy with respect to the product topology. Choueka \cite{Cho74} gave a simple and transparent development of McNaughton's theory, and also studied further the properties and characterizations of the $\omega$-languages recognized by finite state $\omega$-automata. Later, Cohen systematically studied the Chomsky hierarchy for $\omega$-languages by a generalization from classical formal language theory to $\omega$-languages \cite{CG77a,CG77b,CG78b}. Engelfriet studied the generative power of $(\sigma,\rho)$-accepting X-automata on $\omega$-words for any storage type X, where the tuple $(\sigma,\rho)$ defines six acceptance modes \cite{EH93}.

This paper proposes the $(\sigma,\rho,\pi)$-accepting $\omega$-grammar, motivated by the second aspect above. The tuple $(\sigma,\rho,\pi)$ defines various acceptance modes, where $\sigma$ denotes that some designated productions should appear at least once or infinitely often, $\rho$ denotes some binary relation between a set of productions and a designated production set, and $\pi$ denotes normal or leftmost derivations.

In the literature, Cohen only focused on the $\omega$-automata with five types of $i$-acceptance ($i=1$, $1'$, $2$, $2'$, $3$ is the name of acceptance mode) and the $\omega$-grammars with 3-acceptance mode that leads to the Chomsky hierarchy for $\omega$-languages \cite{CG77a}, while Engelfriet studied the $\omega$-automata with six types of $(\sigma,\rho)$-acceptance \cite{EH93}. Since more acceptance modes of $\omega$-automata are considered than $\omega$-grammars, some $(\sigma,\rho)$-accepting $\omega$-automata do not have corresponding models of $\omega$-grammars in the literature.

Therefore, this paper will define $(\sigma,\rho,\pi)$-accepting $\omega$-grammars associated with Engelfriet's $(\sigma,\rho)$-accepting $\omega$-automata. Based on that, we systematically study their relative generative power with respect to $(\sigma,\rho)$-accepting $\omega$-automata.

In particular, we study the equivalence or inclusion relations between $\omega$-grammars and $\omega$-automata by establishing the translation techniques. We will show that, for most of the acceptance modes, the relationship between the two types of $\omega$-devices is similar to the one in the case of finite words. However, for some acceptance modes, the generative power of $\wHyp{CFG}$ is strictly weaker than $\wHyp{PDA}$, and the generative power of $\wHyp{CSG}$ is equal to $\wHyp{TM}$ (rather than linear-bounded $\omega$-automata-like devices). Furthermore, we will raise some remaining open problems for two of the acceptance modes. These open problems show that the relationship between $\omega$-grammars and $\omega$-automata is not easy, although the relationship between grammars and automata on finite words has been well established.

This paper is organized as follows. In Section \ref{Sec:wautomata}, the basic notions of $(\sigma,\rho)$-accepting $\omega$-automata are introduced. In Section \ref{Sec:wgrammar}, the basic notions of $(\sigma,\rho,\pi)$-accepting $\omega$-grammars are formally proposed. In Section \ref{Sec:main_chara}, we recall some known results expressed in terms of our notations. In Sections \ref{Sec:wgrammar_form}, \ref{Sec:wgrammar_lm} and \ref{Sec:wgrammar_nl}, special forms, leftmost derivations and normal derivations of $\omega$-grammar are explored, respectively. In some proofs, we only line out the sketch of the proof, since a formal proof would be quite boring and waste too much space. Finally, related work is discussed in Section \ref{Sec:rwork}, and we conclude in Section \ref{Sec:conc}. Note that a basic knowledge in classical formal language theory \cite{HU79} is assumed in this paper.

\section{$\omega$-Automata and $\omega$-Languages}
\label{Sec:wautomata}

The terminology and notation are mostly taken from \cite{CG77a,CG77b,EH93}, and conform to \cite{HU79}. We may use the terms ``w.r.t.'' and ``s.t.'' denoting ``with respect to'' and ``such that'' respectively.

\begin{definition}
Let $\Sigma$ denote a finite alphabet, $\Sigma^\omega$ denote all infinite ($\omega$-length) strings $u = \prod_{i=1}^\infty a_i$ where $a_i \in \Sigma$. Any member $u$ of $\Sigma^\omega$ is called an \emph{$\omega$-word} or \emph{$\omega$-string}. An \emph{$\omega$-language} is a subset of $\Sigma^\omega$.

For any language $L \subseteq \Sigma^*$ of finite words, define:
\begin{equation*}
    L^\omega = \{ u \in \Sigma^\omega ~|~ u = \prod_{i=1}^\infty x_i \text{, where for each $i$, } \epsilon \neq x_i \in L \}
\end{equation*}
Note that if $L = \{\epsilon\}$ then $L^\omega = \emptyset$. \hfill $\Box$
\end{definition}
In words, $L^\omega$ consists of all $\omega$-words obtained by concatenating words from $L$ in an infinite sequence.

The following definitions will be used to define the acceptance modes for $\omega$-automata and $\omega$-grammars.

\begin{definition}
Let $A$ and $B$ be two sets, for a mapping $f: A \rightarrow B$, we define:
\begin{align*}
\ran(f) &= \{ b ~|~ b\in B, |f^{-1}(b)| \geq 1 \}\\
\inf(f) &= \{ b ~|~ b\in B, |f^{-1}(b)| \text{ is infinite} \}
\end{align*}
where $|X|$ denotes the cardinality of the set $X$. \hfill $\Box$
\end{definition}
In words, the range of values $\ran(f)$ includes the elements in $B$ that appear at least once in the mapping, and $\inf(f)$ includes the elements that appear infinitely many times.

Let $\mathbb{N}$ be the set of natural numbers, $Q$ be a finite set, $f \in Q^\omega$ be an infinite sequence $f = f_1f_2\ldots$. We consider $f$ as a mapping from $\mathbb{N}$ to $Q$ where $f(i) = f_i$. Therefore, $\ran(f)$ is the set of all elements in $Q$ that appear at least once in $f$, and $\inf(f)$ is the set of all elements that appear infinitely often in $f$.

A variety of acceptance modes will now be defined.

\begin{definition}
Let $\sigma:Q^\omega \rightarrow 2^Q$ be a mapping that assigns to each infinite sequence over $Q$ a subset of $Q$,  $\rho$ be a binary relation over $2^Q$, $\mathcal{F} \subseteq 2^Q$ be a set of subsets of $Q$. The infinite sequence $f:\mathbb{N} \rightarrow Q$ is \emph{$(\sigma, \rho)$-accepting w.r.t. $\mathcal{F}$}, if there exists a set $F \in \mathcal{F}$ such that $\sigma(f)\rho F$.  \hfill $\Box$
\end{definition}

As in \cite{EH93}, in the sequel of this paper, we assume that $\sigma \in \{\ran, \inf\}$ and $\rho \in \{\sqcap, \subseteq, =\}$ unless specified, where $A \sqcap B$ means $A \cap B \neq \emptyset$. Thus we will consider the six acceptance modes given in Table \ref{Tab:wAcc}, where the relation between our notation and the five types of $i$-acceptance used in \cite{CG77a} is also included. In the sequel, we will not use the term ``$i$-acceptance'', since the word does not reflect its semantics.

\begin{table}[htb]
\centering
  \begin{tabular}{cccc}
  \hline
  $(\sigma, \rho)$    & $i$-accepting & Semantics & Alias \\
  \hline
  $(\ran, \sqcap   )$ & 1-accepting   & $(\exists F \in \mathcal{F})\ran(f) \cap F \neq \emptyset$ &   \\
  $(\ran, \subseteq)$ & 1'-accepting  & $(\exists F \in \mathcal{F})\ran(f) \subseteq F$ &   \\
  $(\ran, =        )$ &               & $\ran(f) \in \mathcal{F}$ &   \\
  $(\inf, \sqcap   )$ & 2-accepting   & $(\exists F \in \mathcal{F})\inf(f) \cap F \neq \emptyset$ &  \buchi \\
  $(\inf, \subseteq)$ & 2'-accepting  & $(\exists F \in \mathcal{F})\inf(f) \subseteq F$ &   \\
  $(\inf, =        )$ & 3-accepting   & $\inf(f) \in \mathcal{F}$ &  Muller \\
  \hline
  \end{tabular}
  \caption{$f$ is $(\sigma, \rho)$-accepting w.r.t. $\mathcal{F}$}\label{Tab:wAcc}
\end{table}

There exist other types of acceptance mode, such as Rabin's condition. Here we only study these six types, because they are either the most typical ones with applications or more commonly considered in the literature such as \cite{CG77a,CG77b,EH93}. We believe that the results for other acceptance modes can be easily obtained by using similar techniques and methodologies developed in this paper.

The definitions of $\omega$-automata are generalized from those of classical automata by adding a set of designated state sets.

\begin{definition}
A \emph{finite state $\omega$-automaton} ($\wHyp{FSA}$) is a tuple $A = (Q, \Sigma, \delta, q_0, \mathcal{F})$, where $Q$ is a finite set of states, $\Sigma$ is a finite input alphabet, $q_0 \in Q$ is the initial state, $\delta \subseteq Q \times (\Sigma \cup \{\epsilon\}) \times Q$ is a transition function, and $\mathcal{F} \subseteq 2^Q$ is a set of \emph{designated state sets}. If $\delta$ is deterministic, then $A$ is a \emph{deterministic} finite state $\omega$-automaton ($\wHyp{DFSA}$).

Let $u = \prod_{i=1}^\infty a_i \in \Sigma^\omega$, where $\forall i \geq 1$, $a_i \in \Sigma$. A \emph{legal run} (or \emph{complete run}) of $A$ on $u$ is an infinite sequence of states $r = r_1r_2 \ldots$, where $r_1 = q_0$ and $\forall i \geq 1, \exists b_i \in \Sigma \cup \{\epsilon\}$ such that $\delta (r_i,b_i) \ni r_{i+1}$ and $\prod_{i=1}^\infty b_i = \prod_{i=1}^\infty a_i$. \hfill $\Box$
\end{definition}

Note that all computations of $A$ on $u$ which do not correspond to legal runs, e.g. computations which involve infinite $\epsilon$-loops, will be disregarded.

\begin{definition}
A \emph{pushdown $\omega$-automaton} ($\wHyp{PDA}$) is a tuple $D = (Q, \Sigma, \Gamma, \delta, q_0, Z_0, \mathcal{F})$, where $\Gamma$ is a finite stack alphabet, $\delta \subseteq Q \times (\Sigma\cup\{\epsilon\}) \times \Gamma \times Q \times \Gamma^*$ is a transition function, $Z_0 \in \Gamma$ is the start symbol. If $\delta$ is deterministic, then $D$ is a \emph{deterministic} pushdown $\omega$-automaton ($\wHyp{DPDA}$).

A \emph{configuration} of an $\wHyp{PDA}$ is a pair $(q, \gamma)$, where $q \in Q$, $\gamma \in \Gamma^*$ and the leftmost symbol of $\gamma$ is on the top of the stack. For $a \in \Sigma \cup \{\epsilon\}$, $\beta, \gamma \in \Gamma^*$ and $Z \in \Gamma$, we write $a:(q,Z\gamma) \vdash_D (q',\beta\gamma)$ if $\delta(q,a,Z) \ni (q',\beta)$.

Let $u = \prod_{i=1}^\infty a_i \in \Sigma^\omega$, where $\forall i \geq 1$, $a_i \in \Sigma$. A \emph{legal run} (or \emph{complete run}) of $D$ on $u$ is an infinite sequence of configurations $r = \{ (q_i, \gamma_i) \}_{i\geq 1}$, where $(q_1, \gamma_1) = (q_0, Z_0)$ and $\forall i \geq 1, \exists b_i \in \Sigma \cup \{\epsilon\}$ such that  $b_i:(q_i,\gamma_i) \vdash_D (q_{i+1},\gamma_{i+1})$ and $\prod_{i=1}^\infty b_i = \prod_{i=1}^\infty a_i$. \hfill $\Box$
\end{definition}

\begin{definition}
A \emph{Turing $\omega$-machine} ($\wHyp{TM}$) with a single semi-infinite tape is a tuple $M = (Q, \Sigma, \Gamma, \delta$, $q_0, \mathcal{F})$, where $\Gamma$ is a finite tape alphabet such that $\Sigma \subseteq \Gamma$, $\delta \subseteq Q \times \Gamma \times Q \times \Gamma \times \{L,R,S\}$ is a transition function. If $\delta$ is deterministic, then $M$ is a \emph{deterministic} Turing $\omega$-machine ($\wHyp{DTM}$).

A \emph{configuration} of an $\wHyp{TM}$ is a tuple $(q, \gamma, i)$, where $q \in Q$ and $\gamma \in \Gamma^\omega$ and $i \in \mathbb{N}$ indicating the position of the reading head. The relations $\vdash_M$ and $\vdash_M^*$ are defined as usual.

Let $u = \prod_{i=1}^\infty a_i \in \Sigma^\omega$, where $\forall i \geq 1$, $a_i \in \Sigma$. A \emph{run} of $M$ on $u$ is an infinite sequence of configurations $r = \{ (q_i, \gamma_i, j_i) \}_{i\geq 1}$, where $(q_1, \gamma_1, j_1) = (q_0, u, 1)$ and $\forall i \geq 1$, $(q_i,\gamma_i,j_i) \vdash_M (q_{i+1},\gamma_{i+1},j_{i+1})$.

A run $r$ is \emph{complete} if $\forall n \geq 1, \exists k \geq 1$, s.t. $j_k >n$. In words, the whole $\omega$-word will be completely scanned.

A run $r$ is \emph{oscillating} if $\exists n_0 \geq 1, \forall l \geq 1, \exists k \geq l$, s.t. $j_k = n_0$. In words, $n_0$ will be scanned infinitely often.

A \emph{legal run} (or \emph{complete non-oscillating run}, abbreviated c.n.o.) of $M$ on $u$ is a run which is complete and non-oscillating. It corresponds to an infinite computation that scans each square on the tape only finitely many times. \hfill $\Box$
\end{definition}

An \emph{m-tape Turing $\omega$-machine} ($\mwHyp{TM}$) ($m \geq 1$) has $m$ semi-infinite tapes, each with a separate reading head. We assume that initially the input appears on the first tape and the other tapes are blank. The transitions are defined in the usual way \cite{HU79}. The notion of c.n.o. run for an $\mwHyp{TM}$ means an infinite computation that scans each square on \emph{the first tape} only finitely many times. There is no such restriction for the other tapes.

\begin{definition}
A state $q_T \in Q$ is a \emph{traverse state} iff $\forall a \in \Gamma$, $\delta(q_T,a)= \{ (q_T, a, R) \}$. \hfill $\Box$
\end{definition}

The following definitions are common notations for all the types of $\omega$-automata defined above.

\begin{definition}
Let $A$ be an $\omega$-automaton, $u = \prod_{i=1}^\infty a_i \in \Sigma^\omega$, where $\forall i \geq 1$, $a_i \in \Sigma$. Every \emph{legal run} $r$ of $A$ on $u$ induces an infinite sequence of states $f_r = f_1f_2 \ldots$, where $f_1 = q_0$ and $f_i$ is the state entered in the $i$-th step of the legal run $r$. For each $(\sigma, \rho)$, the \emph{$\omega$-language $(\sigma, \rho)$-accepted by $A$} is
\begin{align*}
    L_{\sigma, \rho}(A) = \{ u \in \Sigma^\omega ~|~ &\text{there exists a legal run $r$ of $A$ on $u$ such that}\\
                 &\text{$f_r$ is $(\sigma, \rho)$-accepting w.r.t. $\mathcal{F}$} \}
\end{align*}
Each $\omega$-word $u \in L_{\sigma, \rho}(A)$ is \emph{$(\sigma, \rho)$-accepted\footnote{Sometimes we may also say $(\sigma, \rho)$-generated instead of $(\sigma, \rho)$-accepted, since an $\omega$-automaton can both generate or recognize an $\omega$-word. This also applies to $\omega$-grammars in the sequel.} by} $A$. \hfill $\Box$
\end{definition}

Since $(\inf, =)$-acceptance is the most powerful mode of $\omega$-recognition (i.e. 3-acceptance in \cite{CG77a,CG77b}), we adopt $(\inf, =)$-acceptance as our standard definition of acceptance. Henceforth, $(\inf, =)$-acceptance will be referred to simply as ``acceptance'', and $L_{\inf,=}(A)$ will be denoted by $L(A)$ (the $\omega$-language ``accepted'' by $A$) by omitting $(\inf,=)$.

In the sequel, we denote by $\wHyp{FSA}$, $\wHyp{PDA}$, $\wHyp{TM}$ the families of finite state, pushdown $\omega$-automata, and Turing $\omega$-machines, and denote by $\wHyp{DFSA}$, $\wHyp{DPDA}$, $\wHyp{DTM}$ the families of deterministic ones, respectively. For a family $X$ of $\omega$-automata, we denote the associated family of $(\sigma, \rho)$-accepted $\omega$-languages by $\mathcal{L}_{\sigma, \rho}(X)$. As usual, we denote simply $\mathcal{L}_{\inf, =}(X)$ by $\mathcal{L}(X)$.

\begin{definition}
For each $(\sigma, \rho)$, two $\omega$-automata $A_1$ and $A_2$ are \emph{$(\sigma, \rho)$-equivalent} iff $L_{\sigma, \rho}(A_1) = L_{\sigma, \rho}(A_2)$. They are \emph{equivalent} iff $L(A_1) = L(A_2)$.\hfill $\Box$
\end{definition}

\begin{definition}
An $\omega$-automaton with a unique designated set, i.e., $|\mathcal{F}|=1$, is called a U-$\omega$-automaton. We may denote the unique designated set by $F \subseteq Q$ instead of $\mathcal{F} = \{F\}$. \hfill $\Box$
\end{definition}

\begin{lemma}[Lemma 2.7 of \cite{EH93}]
Let $\sigma \in \{\ran, \inf\}$ and $\rho \in \{\sqcap, \subseteq \}$, for every (deterministic) $\omega$-automaton $A$, there exists a (deterministic) U-$\omega$-automaton $A'$ such that $L_{\sigma, \rho}(A) = L_{\sigma, \rho}(A')$. \hfill $\Box$
\end{lemma}

\begin{definition}
An $\omega$-automaton $A$ has the \emph{continuity property}, abbreviated \emph{Property C}, iff for every $\omega$-words $u \in \Sigma^\omega$ there is a legal run of $A$ on $u$. We say $A$ is a C-$\omega$-automaton. \hfill $\Box$
\end{definition}

Note that the existence of a legal run on $u$ does not necessarily mean that $u$ is accepted, but only means that $u$ does not block the $\omega$-automaton. It is easy to see, by utilizing the nondeterminism, for all $(\sigma, \rho)$-acceptances and $X \in \{\wHyp{FSA},\wHyp{PDA},\wHyp{TM}\}$, every $X$-type $\omega$-automaton $A$ without Property C can be modified into a $(\sigma, \rho)$-equivalent nondeterministic $X$-type $\omega$-automaton $A'$ with Property C.

\section{$\omega$-Grammars and $\omega$-Languages}
\label{Sec:wgrammar}

A \emph{phrase structure grammar} on finite words is denoted $G = (N,T,P,S)$, where $N$ is a finite set of \emph{nonterminals}, $T$ is a finite set of \emph{terminals}, $P$ is a finite set of \emph{productions} of the form $p: \alpha \rightarrow \beta$ where $p$ is the name (or label) of the production, $\alpha \neq \epsilon$, and $\alpha,\beta$ are strings of symbols from $(N \cup T)^*$, and $S \in N$ is the \emph{start symbol}. We define the \emph{vocabulary} $V = N \cup T$. A derivation using a specified production $p$ is denoted by $\alpha \overset{p}\Rightarrow \beta$, and its reflexive and transitive closure is denoted by $\alpha \Rightarrow^* \gamma$, or with the sequence of applied productions $\alpha \overset{p_1...p_k}\Longrightarrow \gamma$. The \emph{language accepted} by $G$ is $L(G) = \{ w \in T^* ~|~ S \Rightarrow^* w\}$.

We denote a leftmost derivation (denoted by $lm$) in the language $L(G)$ by $\alpha \overunderarrow{p_1}{lm} \alpha_1 \overunderarrow{p_2}{lm} \cdots \overunderarrow{p_k}{lm} \alpha_k$. As an abbreviation, we write $\alpha \overunderarrow{p_1...p_k}{lm} \alpha_k$. We will omit ``$lm$'' if there is no confusion.

In this paper, ``leftmost derivation'' means in every step of a derivation, the leftmost nonterminal must be rewritten\footnote{We choose this definition because it is commonly used in the related literature, although there exist other definitions, e.g., only rewriting the leftmost nonterminal that could be rewritten.}. The term ``normal derivation'' means general derivations including those which are leftmost and non-leftmost.

\begin{definition}
A \emph{phrase structure $\omega$-grammar} ($\wHyp{PSG}$) is a quintuple $G = (N, T, P, S, \mathcal{F})$, where $G_1 = (N, T, P, S)$ is an ordinary phrase structure grammar, the productions in $P$ are all of the form $p: \alpha \rightarrow \beta$, where $p$ is the name (or label) of the production, $\alpha \in N^+$, $\beta \in V^*$, and $\mathcal{F} \subseteq 2^P$. The sets in $\mathcal{F}$ are called the \emph{production repetition sets}.

Let $d$ be an infinite derivation in $G$, starting from some string $\alpha \in V^*$:
\begin{equation*}
    d:~\alpha = u_0\alpha_0 \overarrow{p_1} u_0u_1\alpha_1 \overarrow{p_2} \cdots \overarrow{p_i} u_0u_1\cdots u_i\alpha_i \overarrow{p_{i+1}} \cdots
\end{equation*}
where for each $i \geq 0$, $u_{i} \in T^*$, $\alpha_i \in NV^*$, $p_{i+1} \in P$. Note that the derivation need not be leftmost, since some of the $u_i$'s may be empty. We say $d$ is a \emph{leftmost derivation} iff for each $i \geq 1$, the production $p_i$ rewrites the leftmost nonterminal of $\alpha_{i-1}$.

Let $u = \prod_{i=0}^\infty u_i$. If $u \in T^\omega$, we write $d: \alpha \Rightarrow^\omega u$. The assumption that the left-hand side of each production of $P$ is in $N^+$ guarantees that the terminal prefix of each sentential form will never be replaced later in the derivation, and become a prefix of the generated $\omega$-word. The derivation $d$ induces a sequence of productions $d_P = p_1p_2\ldots$, i.e., a mapping $d_P: \mathbb{N} \rightarrow P$ where $d_P(i) = p_i$.

Let $\pi \in \{l,nl\}$ (denoting leftmost and normal derivations, respectively), for each $(\sigma, \rho)$, the \emph{$\omega$-language $(\sigma,\rho,\pi)$-accepted by $G$} is
\begin{align*}
    L_{\sigma,\rho,l}(G) = \{u \in T^\omega ~|~ &\text{there exists a leftmost derivation $d:S \overunderarrow{}{lm}^\omega u$ in $G$}\\
    &\text{such that $d_P$ is $(\sigma,\rho)$-accepting w.r.t. $\mathcal{F}$} \} \\
    L_{\sigma,\rho,nl}(G) = \{u \in T^\omega ~|~ &\text{there exists a derivation $d:S \Rightarrow^\omega u$ in $G$}\\
    &\text{such that $d_P$ is $(\sigma,\rho)$-accepting w.r.t. $\mathcal{F}$} \}
\end{align*}
As usual, $L_{\inf,=,\pi}(G)$ will be denoted by $L_\pi(G)$. \hfill $\Box$
\end{definition}

\begin{definition}
A \emph{context sensitive $\omega$-grammar} ($\wHyp{CSG}$) is an $\wHyp{PSG}$ in which for each production $\alpha \rightarrow \beta$, $|\beta| \geq |\alpha|$ holds. \hfill $\Box$
\end{definition}
This type of grammar is also called monotonic or length-increasing grammar. In order to keep conformance with the literature, we choose it as the definition of $\wHyp{CSG}$.

\begin{definition}
A \emph{context-free $\omega$-grammar ($\wHyp{CFG}$) with production repetition sets} is an $\wHyp{PSG}$ whose productions are of the form $A \rightarrow \alpha$, $A \in N$, $\alpha \in (N \cup T)^*$. \hfill $\Box$
\end{definition}

\begin{definition}
A \emph{right linear $\omega$-grammar ($\wHyp{RLG}$) with production repetition sets} is an $\wHyp{PSG}$ whose productions are of the form $A \rightarrow uB$ or $A \rightarrow u$, $A,B \in N$, $u \in T^*$. \hfill $\Box$
\end{definition}

In the sequel, we denote by $\wHyp{RLG}$, $\wHyp{CFG}$, $\wHyp{CSG}$, $\wHyp{PSG}$ the families of right-linear, context-free, context-sensitive, arbitrary phrase structure $\omega$-grammars, respectively. For a family $X$ of $\omega$-grammars, we denote the associated families of $(\sigma,\rho,\pi)$-accepted $\omega$-languages by $\mathcal{L}_{\sigma, \rho, \pi}(X)$. As we mentioned, we denote simply $\mathcal{L}_{\inf, =, \pi}(X)$ by $\mathcal{L}_{\pi}(X)$.

\begin{definition}
For each $(\sigma, \rho,\pi)$, two $\omega$-grammars $G_1$ and $G_2$ are \emph{$(\sigma,\rho,\pi)$-equivalent} iff $L_{\sigma,\rho,\pi}(G_1) = L_{\sigma, \rho,\pi}(G_2)$. They are \emph{equivalent in $\pi$-derivation} iff $L_\pi(G_1) = L_\pi(G_2)$.\hfill $\Box$
\end{definition}

\begin{definition}
An $\omega$-grammar with a unique designated set, i.e., $|\mathcal{F}|=1$, is called a U-$\omega$-grammar. We may denote the unique designated set by $F \subseteq P$ instead of $\mathcal{F} = \{F\}$. \hfill $\Box$
\end{definition}

\begin{definition}
An $\omega$-grammar with a designated set $\mathcal{F}=2^P$ is a \emph{unrestricted $\omega$-grammar}, denoted by u-$\omega$-grammar. \hfill $\Box$
\end{definition}

The previous definitions concern the $\omega$-grammars with production repetitions sets. Now we switch to the $(\sigma,\rho,\pi)$-acceptance w.r.t. variable repetition sets of context-free $\omega$-grammars.

\begin{definition}
A \emph{context-free $\omega$-grammar with variable repetition sets} ($\wHypV{CFG}$) is a quintuple $G = (N, T, P, S, \mathcal{F})$, where $G_1 = (N, T, P, S)$ is an ordinary context-free grammar and $\mathcal{F} \subseteq 2^N$. The sets in $\mathcal{F}$ are called the \emph{variable repetition sets}.

Let $d: \alpha \Rightarrow^\omega u \in T^\omega$ be an infinite derivation in $G$. The derivation $d$ induces a sequence of nonterminals $d_N = n_1n_2\ldots$, i.e., a mapping $d_N: \mathbb{N} \rightarrow N$ with $d_N(i) = n_i$, where $n_i \in N$ is the nonterminal which is the left-hand side of the $i$-th production in $d_P$.

Let $\pi \in \{l,nl\}$, for each $(\sigma, \rho)$, the \emph{$\omega$-language $(\sigma,\rho,\pi)$-accepted by $G$} is
\begin{align*}
    L_{\sigma,\rho,l}(G) = \{u \in T^\omega ~|~ &\text{there exists a leftmost derivation $d:S \overunderarrow{}{lm}^\omega u$ in $G$}\\
    &\text{such that $d_N$ is $(\sigma,\rho)$-accepting w.r.t. $\mathcal{F}$} \} \\
    L_{\sigma,\rho,nl}(G) = \{u \in T^\omega ~|~ &\text{there exists a derivation $d:S \Rightarrow^\omega u$ in $G$}\\
    &\text{such that $d_N$ is $(\sigma,\rho)$-accepting w.r.t. $\mathcal{F}$} \}
\end{align*}
As usual, $L_{\inf,=,\pi}(G)$ will be denoted by $L_\pi(G)$.\hfill $\Box$
\end{definition}

\begin{definition}
A \emph{right linear $\omega$-grammar with variable repetition sets} ($\wHypV{RLG}$) is an $\wHypV{CFG}$ whose productions are of the form $A \rightarrow uB$ or $A \rightarrow u$, $A,B \in N$, $u \in T^*$. \hfill $\Box$
\end{definition}

The following theorem states that the $(\inf,=,\pi)$-acceptance w.r.t. the two types of repetition sets defined above are equivalent in generative power. The proofs of the two equations can be found in Remark 2.7 and Proposition 4.1.1 of \cite{CG77b}.
\begin{theorem}[Thm. 3.1.4 of \cite{CG77a}]\label{Thm:wCFG_wCFGV}
(1) $\mathcal{L}_l(\wHyp{CFG}) = \mathcal{L}_l(\wHypV{CFG})$.\\
(2) $\mathcal{L}_{nl}(\wHyp{CFG}) = \mathcal{L}_{nl}(\wHypV{CFG})$. \hfill $\Box$
\end{theorem}

Note that for right linear $\omega$-grammars, every derivation is a leftmost derivation. Thus we have the following theorem.
\begin{theorem}
$\mathcal{L}_l(\wHyp{RLG}) = \mathcal{L}_{nl}(\wHyp{RLG}) = \mathcal{L}_l(\wHypV{RLG}) = \mathcal{L}_{nl}(\wHypV{RLG})$.
\end{theorem}

However, for the $(\inf,=)$-acceptance mode, the leftmost generation of $\wHyp{CFG}$ is strictly more powerful than its normal generation.
\begin{theorem}[Thm. 4.3.7 of \cite{CG77b}]
(1) $\mathcal{L}_{nl}(\wHyp{CFG}) \subset \mathcal{L}_l(\wHyp{CFG})$.\\
(2) $\mathcal{L}_{nl}(\wHypV{CFG}) \subset \mathcal{L}_l(\wHypV{CFG})$. \hfill $\Box$
\end{theorem}

Therefore, we choose the leftmost derivation as our standard definition of acceptance in $\wHyp{CFG}$'s. That means, $L_{\sigma,\rho,l}(G)$, $L_{l}(G)$, $\mathcal{L}_{\sigma,\rho,l}(X)$, $\mathcal{L}_{l}(X)$ will be denoted simply by $L_{\sigma,\rho}(G)$, $L(G)$, $\mathcal{L}_{\sigma,\rho}(X)$, $\mathcal{L}(X)$, respectively.

\section{Main Characterizations}
\label{Sec:main_chara}

In this section, we recall some known results expressed in terms of our notation. These results constitute the Chomsky hierarchy of $\omega$-languages.

We denote by $REGL$, $CFL$, $REL$ the families of regular, context-free, recursive enumerable languages of finite words, and denote by $\wHyp{REGL}$, $\wHyp{CFL}$, $\wHyp{REL}$ the families of $\omega$-type ones that will be defined in this section, respectively.

\begin{definition}
For any family $\mathcal{L}$ of languages of finite words over alphabet $\Sigma$, the \emph{$\omega$-Kleene closure} of $\mathcal{L}$, denoted by $\wHyp{KC}(\mathcal{L})$, is:
\begin{equation*}
    \wHyp{KC}(\mathcal{L}) = \{ L\subseteq \Sigma^\omega ~|~ L= \bigcup_{i=1}^k U_i V_i^\omega \text{ for some $U_i$, $V_i \in \mathcal{L}$, $1 \leq i \leq k$, and $k \in \mathbb{N}$} \}
\end{equation*}
where $\mathbb{N}$ is the set of natural numbers. \hfill $\Box$
\end{definition}

The main characterization theorem for regular $\omega$-languages is the following one.
\begin{theorem}[Thm. 2.2.2 and 3.1.9 of \cite{CG77a}, \cite{Buc60}, \cite{McN66}]
For any $\omega$-language $L \subseteq \Sigma^\omega$, the following conditions are equivalent:
\begin{enumerate}
  \item $L \in \wHyp{KC}(REGL)$
  \item $L \in \mathcal{L}_{\inf,=}(\wHyp{FSA})$
  \item $L \in \mathcal{L}_{\inf,\sqcap}(\wHyp{FSA})$
  \item $L \in \mathcal{L}_{\inf,=}(\wHyp{DFSA})$
  \item $L \in \mathcal{L}_{\inf,=,l}(\wHypV{RLG})$
\end{enumerate}
The $\omega$-language $L$ is a \emph{regular $\omega$-language} ($\wHyp{REGL}$), if it satisfies the conditions. It is \emph{effectively given} if it is given in one of the forms above. \hfill $\Box$
\end{theorem}
Note that, in Thm. 2.2.2 of \cite{CG77a}, Item (3) was $L \in \mathcal{L}_{\inf,\sqcap}(\HypwHyp{U}{FSA})$. Here we provide a more generic result, since it is easy to show $\mathcal{L}_{\inf,\sqcap}(\HypwHyp{U}{FSA}) = \mathcal{L}_{\inf,\sqcap}(\wHyp{FSA})$ by combining the designated state sets.

\begin{theorem}[Thm. 2.2.4 of \cite{CG77a}, Thm. 1.8 and 1.12 of \cite{CG77b}, \cite{Buc60}, \cite{McN66}]
The family of regular $\omega$-languages ($\wHyp{REGL}$, i.e., $\mathcal{L}(\wHyp{FSA})$) is closed under all Boolean operations, regular substitution and generalized sequential machine (gsm) mapping.\hfill $\Box$
\end{theorem}

\begin{theorem}[Thm. 2.2.5 of \cite{CG77a}]
For any regular $\omega$-languages $L_1$ and $L_2$ effectively given, it is decidable whether (1) $L_1$ is empty, finite or infinite; (2) $L_1 = L_2$; (3) $L_1 \subseteq L_2$; (4) $L_1 \cap L_2 = \emptyset$. \hfill $\Box$
\end{theorem}

The main characterization theorem for context-free $\omega$-languages is the following one.
\begin{theorem}[Thm. 4.1.8 of \cite{CG77a}]\label{Thm:wCFL_mc}
For any $\omega$-language $L \subseteq \Sigma^\omega$, the following conditions are equivalent:
\begin{enumerate}
  \item $L \in \wHyp{KC}(CFL)$
  \item $L \in \mathcal{L}_{\inf,=}(\wHyp{PDA})$
  \item $L \in \mathcal{L}_{\inf,\sqcap}(\wHyp{PDA})$
  \item $L \in \mathcal{L}_{\inf,=,l}(\wHypV{CFG})$
\end{enumerate}
The $\omega$-language $L$ is a \emph{context-free $\omega$-language} ($\wHyp{CFL}$), if it satisfies the conditions. It is \emph{effectively given} if it is given in one of the forms above. \hfill $\Box$
\end{theorem}

\begin{theorem}[Section 1 of \cite{CG77b}]
The family of context-free $\omega$-languages ($\wHyp{CFL}$, i.e., $\mathcal{L}(\wHyp{PDA})$) is closed under union, intersection with $\wHyp{REGL}$, quotient with $\wHyp{REGL}$, context-free substitution and gsm mapping, is not closed under intersection and complementation.\hfill $\Box$
\end{theorem}

\begin{theorem}[Thm. 4.2.6 and 4.2.8 of \cite{CG77a}]
For any context-free $\omega$-language $L$ and regular $\omega$-language $R$ effectively given, it is decidable whether (1) $L$ is empty, finite or infinite; (2) $L \subseteq R$. \hfill $\Box$
\end{theorem}

Before we present the main characterization theorem for recursive enumerable $\omega$-languages, we would like to first prove a theorem about multi-tape $\mwHyp{TM}$.

For completeness, we now define a folding process of Turing machine, which will enable us to turn every complete run into a c.n.o. run (see Section 6 of \cite{CG78b}).

\begin{figure*}[bth]
    \centering
    \includegraphics[scale=1]{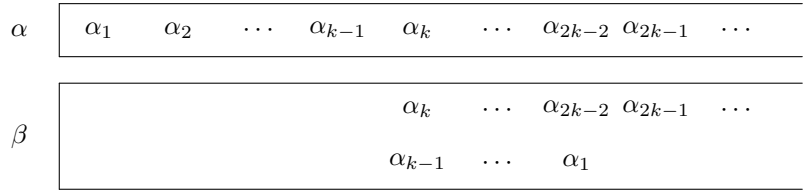}\\
    \caption{$\beta$ is a $k$-folded version of $\alpha$}\label{Fig:kFolded}
\end{figure*}

\begin{definition}[$k$-Folded Version]
Let $\alpha$, $\beta$ be infinite tapes, where $\beta$ has two tracks. We say $\beta$ is a \emph{$k$-folded version} of $\alpha$, iff:
\begin{enumerate}
  \item for $k \leq j \leq 2k-2$, $\beta_j$ contains $\alpha_j$ on its first track and $\alpha_{2k-j-1}$ on its second track.
  \item for $j > 2k-2$, $\beta_j$ contains $\alpha_j$ on its first track. \hfill $\Box$
\end{enumerate}
\end{definition}
In words, as shown in Fig. \ref{Fig:kFolded}, the $\omega$-word on $\alpha$ is divided into two parts $\alpha_1\alpha_2\cdots\alpha_{k-1}$ and $\alpha_k\alpha_{k+1}\cdots$, and the first part is folded forwards on the second track of $\beta$, while the second part is placed on the first track of $\beta$ and at the same position as on $\alpha$.

\begin{definition}[Relative Folding Process]
Let $M$ be an $\mwHyp{TM}$ and $\alpha,\beta$ be two working tapes (or two tracks of a single tape). We can construct an $\mwHyp{TM}$ $M_1$ by applying the \emph{relative folding process of $\beta$ w.r.t. $\alpha$}: $M_1$ simulates $M$ on $\omega$-input $u$, for each $i \geq 2$, whenever $M_1$ scans $\alpha_i$ for the first time, $M_1$ will create the $i$-folded version of the ($i$-1)-folded $\omega$-word on $\beta$, and then it will continue the simulation. \hfill $\Box$
\end{definition}

\begin{lemma}
Let $M$ be an $\mwHyp{TM}$ and $\alpha,\beta$ be two working tapes (or two tracks of a single tape). For every $(\sigma, \rho)$, there can be constructed a $(\sigma,\rho)$-equivalent $\mwHyp{TM}$ $M_1$ by applying \emph{relative folding process} to $\beta$ w.r.t. $\alpha$ with the following property: $M_1$ simulates $M$ on an $\omega$-input $u$ such that, for each $i \geq 2$, within some finite computation steps after $\alpha_i$ has been reached for the first time, $M_1$'s reading head on $\beta$ will be to the right of $\beta_{i-1}$ and will never again return to the initial segment $\beta_1\cdots\beta_{i-1}$.
\end{lemma}
\proof In $M_1$, we may add a boolean component to the state of $M$, denoting the head on $\beta$ is on the first track or on the second track. Whenever $M_1$ scans $\alpha_i$ for the first time, $M_1$ will create the $i$-folded version of the ($i$-1)-folded $\omega$-word on $\beta$, within some finite computation steps. After that, whenever $M$ tries to access the initial segment $\beta_1\cdots\beta_{i-1}$, $M_1$ can simulate by moving the head on $\beta$ to the second track (positions $k$ to $2k-2$) to access the content of the segment, but without actually returning to the initial segment. \hfill $\Box$

\begin{theorem}\label{Thm:mwTM_equ_wTM}
For every $\mwHyp{TM}$, $m \geq 1$, there can be constructed a $(\sigma,\rho)$-equivalent $\wHyp{TM}$, for every $(\sigma, \rho)$. Therefore, $\mathcal{L}_{\sigma,\rho}(\mwHyp{TM}) = \mathcal{L}_{\sigma,\rho}(\wHyp{TM})$, for every $(\sigma, \rho)$.
\end{theorem}
\proof The proof resembles that of Thm. 7.3 of \cite{CG78b}. If $m \geq 3$, then $\mwHyp{TM}$ can be translated into a $\HypwHyp{2}{TM}$ of which the second tape simulates the former's $m-1$ working tapes by using $m-1$ tracks. For every $\HypwHyp{2}{TM}$ $M$ with the set of designated sets $\mathcal{F}$, there can be constructed an $\wHyp{TM}$ $M'$ which simulates $M$ by two tracks $\alpha$, $\beta$ on the tape. The two tracks are used to simulate the two tapes of $M$, respectively. The simulation applies the relative folding process to $\beta$ w.r.t. $\alpha$. This will guarantee that every c.n.o. run of $M$ is simulated by a c.n.o. run of $M'$. For each $(\sigma,\rho)$-acceptance, one can define a set of designated sets $\mathcal{H}$ to finish the proof. \hfill $\Box$

\begin{theorem}[Theorems 5.1, 5.9 and 8.2 of \cite{CG78b}]
For any $\omega$-language $L \subseteq \Sigma^\omega$, the following conditions are equivalent:
\begin{enumerate}
  \item $L \in \mathcal{L}_{\sigma,\rho}(\wHyp{TM})$, for $\sigma \in \{\ran, \inf\}$ and $\rho \in \{\sqcap, \subseteq, =\}$
  \item $L \in \mathcal{L}_{\sigma,\rho}(\mwHyp{TM})$, for $\sigma \in \{\ran, \inf\}$ and $\rho \in \{\sqcap, \subseteq, =\}$
  \item $L \in \mathcal{L}_{\inf,=,nl}(\wHyp{PSG})$
  \item $L \in \mathcal{L}_{\inf,=,nl}(\wHyp{CSG})$
\end{enumerate}
The $\omega$-language $L$ is a \emph{recursive enumerable $\omega$-language} ($\wHyp{REL}$), if it satisfies the conditions. Note that $\wHyp{KC}(REL) \subset \wHyp{REL}$. \hfill $\Box$
\end{theorem}
Note that Item (1) extends Thm. 8.2 of \cite{CG78b} where only $i$-acceptances are considered, and Item (2) follows from Thm. \ref{Thm:mwTM_equ_wTM}.

\begin{theorem}[Section 5.3 and Thm. 8.4 of \cite{CG78b}]
The family of recursive enumerable $\omega$-languages ($\wHyp{REL}$, i.e., $\mathcal{L}(\wHyp{TM})$) is closed under union, intersection, recursive enumerable substitution and concatenation with recursive enumerable languages, is not closed under complementation. \hfill $\Box$
\end{theorem}

The following result shows inclusion or equivalence between the families of $\omega$-languages recognized by various $(\sigma,\rho)$-accepting $X$-type $\omega$-automata.
\begin{theorem}[Thm. 3.5 of \cite{EH93}]\label{Thm:Gen_X_wAutomata}
For the various $(\sigma,\rho)$-acceptance modes of X-type $\omega$-automata, $X \in \{\wHyp{FSA},\wHyp{PDA},\wHyp{TM}\}$, we have $\mathcal{L}_{\ran,\subseteq}(X) \subseteq \mathcal{L}_{\ran,\sqcap}(X) = \mathcal{L}_{\ran,=}(X) = \mathcal{L}_{\inf,\subseteq}(X) \subseteq \mathcal{L}_{\inf,\sqcap}(X) = \mathcal{L}_{\inf,=}(X)$. \hfill $\Box$
\end{theorem}
Note that this is only a generic result. For some specific type of X-automata, the inclusions may be strict, e.g. for $\wHyp{PDA}$ \cite{CG77b}.

\section{Special Forms of $\omega$-Grammar}
\label{Sec:wgrammar_form}
The rest of this paper is devoted to explore the relationships between $(\sigma,\rho,\pi)$-accepting $\omega$-grammars and $(\sigma,\rho)$-accepting $\omega$-automata. The results hold for all $(\sigma,\rho,\pi)$-acceptance modes unless explicitly specified. Before studying leftmost and normal derivations of $\omega$-grammar, we would like to introduce some special forms of $\omega$-grammar that will be used to study the generative power in the sequel. Some of the results are generalized from the grammars on finite words by taking into account the production repetition sets.

We start with a special form of $\wHyp{RLG}$. This form guarantees that there is at most one terminal on the right-hand side of a production.

\begin{lemma}\label{Lem:wRLG_no_Tstar_p}
Given an $\wHyp{RLG}$ $G = (N,T,P,S_0,\mathcal{F})$, there can be constructed a $(\sigma,\rho,\pi)$-equivalent $\wHyp{RLG}$ $G' = (N',T,P',S_0,\mathcal{H})$ whose productions are of the form $A \rightarrow aB$ or $A \rightarrow a$, $a \in T \cup \{\epsilon\}$, such that $L_{\sigma,\rho,\pi}(G) = L_{\sigma,\rho,\pi}(G')$, for every $(\sigma, \rho,\pi)$.
\end{lemma}
\proof Without loss of generality, assume that $N = \{S_k\}_{0 \leq k < |N|}$, $P = \{p_k\}_{1 \leq k \leq |P|}$ consists of all productions of the forms: $p_k: S_i \rightarrow u S_j$, $p_k: S_i \rightarrow u$, $u \in T^*$. We construct $G'$ as follows: $P'$ consists of all productions of the following forms:
\begin{enumerate}
  \item $p_k: S_i \rightarrow a S_j$, if $p_k: S_i \rightarrow a S_j \in P$, $a \in T \cup \{\epsilon\}$.
  \item $p_k: S_i \rightarrow a$, if $p_k: S_i \rightarrow a \in P$, $a \in T \cup \{\epsilon\}$.
  \item $p_{km}: S_{k(m-1)} \rightarrow a_{km} S_{km}$, for $1 \leq m \leq |u|$, where $S_{k0} = S_i$, $S_{k|u|} = S_j$, if $p_k: S_i \rightarrow u S_j \in P$, where $u = a_{k1} \cdots a_{k|u|}$, $|u| \geq 2$.
  \item $p_{km}: S_{k(m-1)} \rightarrow a_{km} S_{km}$, for $1 \leq m \leq |u|$, where $S_{k0} = S_i$, $S_{k|u|} = \epsilon$, if $p_k: S_i \rightarrow u \in P$, where $u = a_{k1} \cdots a_{k|u|}$, $|u| \geq 2$.
\end{enumerate}
In words, Items (3) and (4) decompose each long production into several short productions, and each new intermediate nonterminal $S_{km}$ is unique in $N'$.

We denote by $P_k$ the set of productions named $p_k$ or $p_{km}$. Thus, the original production $p_k$ is simulated by the set $P_k$. Let $\mathcal{F} = \{F_i\}_{1\leq i \leq n}$, we construct the set $\mathcal{H} = \{ H_i\}_{1\leq i \leq n}$ where $H_i = \bigcup_{p_k \in F_i} P_k$. It can be easily verified that $L_{\sigma,\rho,\pi}(G) = L_{\sigma,\rho,\pi}(G')$ for all $(\sigma,\rho,\pi)$-acceptance modes. \hfill$\Box$

The next lemma concerns the $\epsilon$-production-free $\wHyp{CFG}$. This form does not include $\epsilon$-productions, which guarantees that an $\wHyp{CFG}$ is also an $\wHyp{CSG}$, and that the length of sentential form will never decrease in a derivation.

\begin{lemma}\label{Lem:wCFG_no_epsilon}
Given an $\wHyp{CFG}$ $G = (N,T,P,S,\mathcal{F})$, there can be constructed a $(\sigma,\rho,\pi)$-equivalent $\epsilon$-production-free $\wHyp{CFG}$ $G' = (N',T,P',S,\mathcal{H})$ with no productions of the form $A \rightarrow \epsilon$, such that $L_{\sigma,\rho,\pi}(G) = L_{\sigma,\rho,\pi}(G')$, for $(\sigma,\rho,\pi) \not\in \{(\ran, \sqcap, l),(\ran, =, l)\}$.
\end{lemma}
\proof Define $NL(\alpha) = \{ D \subseteq P ~|~$ there exists a finite derivation $d: \alpha \Rightarrow^* \epsilon$ s.t. $\ran(d_P) = D\}$, i.e. the set of production sets that can rewrite $\alpha$ to a null string. Define the substitution $h$ as: for $A \in N$, $h(A) = A$ if $NL(A) = \emptyset$, $h(A) = \{A,\epsilon\}$ if $NL(A) \neq \emptyset$, and for $a \in T$, $h(a) =a$.

Let $\alpha = \prod_{i=1}^l A_i$ be a sentential form, and $\beta = \prod_{i=1}^l B_i \in h(\alpha)$, where $\forall 1 \leq i \leq l$, $A_i \in N \cup T$, $B_i \in h(A_i)$. To accumulate the productions that are applied to obtain $\beta$ by rewriting some nonterminals in $\alpha$ to be $\epsilon$, we define $PE(\beta) = \{ \bigcup_{i=1}^l P_i ~|~ $ for $1 \leq i \leq l$, $P_i = \emptyset$ if $B_i =A_i$, or $P_i \in NL(A_i)$ if $B_i = \epsilon \}$.

Now we construct $G'$. Let $P' = \{[p,K,\beta]: A \rightarrow \beta ~|~ p: A\rightarrow \alpha \in P, \epsilon \neq \beta \in h(\alpha), K \in PE(\beta)\} $. Define $Pro([p,K,\beta]) = \{p\} \cup K$ for each production in $P'$, and $Pro(H) = \bigcup_{p\in H} Pro(p)$ for a set $H$ of productions. In words, $Pro$ accumulates to a single set all the productions that simulate together the derivation $A \Rightarrow \alpha \Rightarrow^* \beta$.

\begin{figure*}[bth]
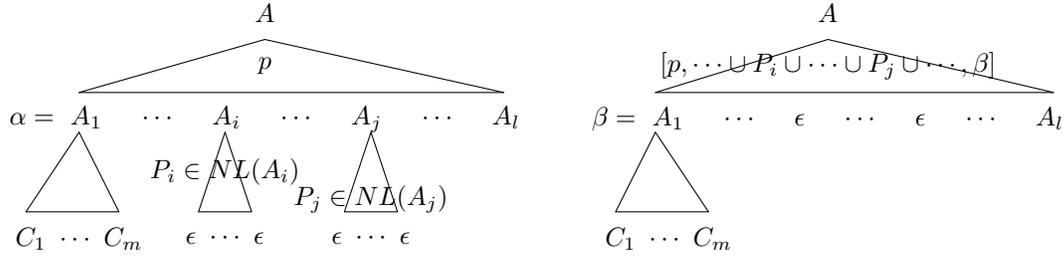

    \centering
    \begin{minipage}[t]{.45\textwidth}
        \centering
        \includegraphics[scale=1]{fig_wCFG_e_02.mps}\\
    \end{minipage}~~
    \begin{minipage}[t]{.45\textwidth}
        \centering
        \includegraphics[scale=1]{fig_wCFG_e_03.mps}\\
    \end{minipage}
    \caption{Simulation by an $\epsilon$-production-free $\wHyp{CFG}$}\label{Fig:wCFG_no_epsilon}
\end{figure*}

An intuitive view of the simulation is shown in Fig. \ref{Fig:wCFG_no_epsilon}. The left figure shows a derivation of $G$. The production $p$ rewrites $A$ to be $\alpha$, then some nonterminals in $\alpha$ (such as $A_i,A_j$) are further rewritten to be $\epsilon$. The right figure shows a simulating derivation of $G'$: $A$ is directly rewritten to be $\beta$, where $A_i,A_j$ are replaced by $\epsilon$. Note that in the simulation without $\epsilon$-productions, the productions that rewrite $A_i$ and $A_j$ to be $\epsilon$ are accumulated in the name of the production, i.e. the second component $K \in PE(\beta)$.

Let $\mathcal{F} = \{F_k\}_{1\leq k \leq n}$, we construct the set $\mathcal{H}$ according to different acceptance modes:
\begin{enumerate}
\item $(\ran,\sqcap,nl)$-acceptance. $\mathcal{H} = \{\{ p \in P' ~|~ Pro(p) \cap \bigcup_{k=1}^n F_k \neq \emptyset \}\}$.
\item $(\ran,\subseteq,nl)$-acceptance. Let $\mathcal{H}_k = \{ H \subseteq P' ~|~ Pro(H) \subseteq F_k \}$, then $\mathcal{H} = \bigcup_{1\leq k \leq n} \mathcal{H}_k$.
\item $(\ran,=,nl)$-acceptance. $\mathcal{H} = \{ H \subseteq P' ~|~ Pro(H) \in \mathcal{F} \}$.
\item $(\inf,\sqcap,nl)$-acceptance. The same as (1).
\item $(\inf,\subseteq,nl)$-acceptance. The same as (2).
\item $(\inf,=,nl)$-acceptance. The same as (3).
\item $(\ran,\subseteq,l)$-acceptance. The same as (2).
\item $(\inf,\subseteq,l)$-acceptance. The same as (2).
\end{enumerate}
For the above cases, it can be easily verified that $L_{\sigma,\rho,\pi}(G) = L_{\sigma,\rho,\pi}(G')$. However, the constructive proof does not work for other four acceptance modes.

Now we consider $(\inf, =, l)$-acceptance. We have $\mathcal{L}_{\inf, =, l}(\wHyp{CFG}) = \wHyp{KC}(CFL)$ by Theorems \ref{Thm:wCFL_mc} and \ref{Thm:wCFG_wCFGV}. Therefore, every $\omega$-language $L$ that is $(\inf,=,l)$-accepted by $G$ can be expressed in the form $L = \bigcup_{i=1}^k U_i V_i^\omega$, for some natural number $k$, and $\forall 1\leq i \leq k$, $\epsilon \not\in U_i \in CFL$, $\epsilon \not\in V_i \in CFL$. Obviously, the $2k$ context-free languages $\{U_i,V_i\}_{1\leq i \leq k}$ can be generated by $2k$ $\epsilon$-production-free context-free grammars $\{G_i,G'_i\}_{1\leq i \leq k}$ respectively, such that $L = \bigcup_{i=1}^k L(G_i) L(G'_i)^\omega$. Thus, one can easily construct an $\epsilon$-production-free $\wHyp{CFG}$ $G'$ that accepts $L$ from the $2k$ $\epsilon$-production-free context-free grammars.

Now we consider $(\inf, \sqcap, l)$-acceptance. It is easy to show $\mathcal{L}_{\inf,\sqcap,l}(\wHyp{CFG}) \subseteq \mathcal{L}_{\inf,=,l}(\wHyp{CFG}) = \wHyp{KC}(CFL) \subseteq \mathcal{L}_{\inf,\sqcap,l}(\wHyp{CFG})$. Thus $\mathcal{L}_{\inf,\sqcap,l}(\wHyp{CFG}) = \wHyp{KC}(CFL)$. The construction of $G'$ is similar to $(\inf,=,l)$-acceptance. \hfill$\Box$

Unfortunately, for $(\ran, \sqcap, l)$-acceptance and $(\ran, =, l)$-acceptance, the construction of $\epsilon$-production-free $\wHyp{CFG}$ is still an open problem. If we use a similar construction as the $nl$-derivation case, the difficulty comes from how to simulate any derivation of $G'$ by using $G$: $G'$ may apply a production $[p,K,\beta]:A \rightarrow \beta$ in $P'$, where some $B_i$ in $\beta$ is supposed to be $\epsilon$ ($A_i$ is rewritten by some productions to be $\epsilon$), but $A_i$ cannot be reached by the corresponding leftmost derivation of the original $\omega$-grammar $G$.

The following lemma concerns a normal form of $\wHyp{PSG}$. This form guarantees that terminals only appear in the right-hand sides of the productions of the form $A \rightarrow a$.

\begin{lemma}\label{Lem:wPSG_NF_1}
Given an $\wHyp{PSG}$ $G = (N,T,P,S,\mathcal{F})$, there can be constructed a $(\sigma,\rho,\pi)$-equivalent $\wHyp{PSG}$ $G' = (N',T,P',S,\mathcal{H})$ whose productions are of the forms $\alpha \rightarrow \beta$, $A \rightarrow a$ or $A \rightarrow \epsilon$, $\alpha,\beta \in N^+$, $A \in N$, $a \in T$, such that $L_{\sigma,\rho,\pi}(G) = L_{\sigma,\rho,\pi}(G')$, for every $(\sigma, \rho,\pi)$.
\end{lemma}
\proof Without loss of generality, assume that $P = \{p_k\}_{1 \leq k \leq |P|}$, and the maximal length of the right-hand sides of the productions is $l$. We construct $G'$ with $N' = N \cup \{ b_{ki} ~|~p_k \in P, 1 \leq i \leq l\} \cup \{E_k ~|~ p_k \in P\}$. $P'$ consists of all productions of the following forms:
\begin{enumerate}
  \item $p_k: \alpha \rightarrow \beta$, if $p_k: \alpha \rightarrow \beta \in P$, $\alpha,\beta \in N^+$.
  \item $p_k: A \rightarrow \epsilon$, if $p_k: A \rightarrow \epsilon \in P$, $A \in N$.
  \item $p_k: \alpha \rightarrow b_{k1}...b_{k|\gamma|}$, if $p_k: \alpha \rightarrow \gamma \in P$, $\alpha \in N^+$, $\gamma = a_1...a_{|\gamma|} \in V^+ -N^+$, where $b_{ki} = a_i$ for each $a_i \in N$, and $p_{ki}: b_{ki} \rightarrow a_i$ for each $a_i \in T$.
  \item $p_k: \alpha \rightarrow E_k$ and $p_{k\epsilon}: E_k \rightarrow \epsilon$, if $p_k: \alpha \rightarrow \epsilon \in P$, $\alpha \in N^+ - N$.
\end{enumerate}
In Item (3), $a_{ki}$ is used to replace the terminal $a_i$ at the position $i$ of production $p_k$ by a nonterminal, and guarantee the nonterminal is unique in the set $N'$. Similarly, in Item (4), $E_k$ is used to replace $\epsilon$ in an $\epsilon$-production $p_k$.

We define the function $f$ as follows:
            \begin{equation*}
              f(p_k) = \left\{ \begin{aligned}
                          &\{p_k\}, \text{ if $p_k \in P$ is of the form in case (1) or (2). } \\
                          &\{p_k\} \cup \{p_{ki} ~|~ a_i \in T \text{ is the $i$-th symbol of $\gamma$} \},\\ &~~~~~~~~~~~~~~~~\text{ if $p_k \in P$ is of the form in case (3). } \\
                          &\{p_k,p_{k\epsilon}\}, \text{ if $p_k \in P$ is of the form in case (4). } \\
                          \end{aligned} \right.
            \end{equation*}
For a set $H$ of productions, $f(H) = \bigcup_{p_k \in H} f(p_k)$. Let $\mathcal{F} = \{F_i\}_{1\leq i \leq n}$, we construct the set $\mathcal{H}_i = \{ H ~|~F_i \subseteq H \subseteq f(F_i) \}$, then $\mathcal{H} = \bigcup_{1\leq i \leq n} \mathcal{H}_i$. In the bisimulation of $G$ and $G'$, $G$ uses production $p_k$, iff $G'$ uses a set of productions $R$ such that $F_i \subseteq R \subseteq f(F_i)$. It can be easily verified that $L_{\sigma,\rho,\pi}(G) = L_{\sigma,\rho,\pi}(G')$ for all $(\sigma,\rho,\pi)$-acceptance modes. \hfill$\Box$

We have similar lemmas for $\wHyp{CFG}$ and $\wHyp{CSG}$, by using the same proof technique as the above one.
\begin{lemma}\label{Lem:wCFG_NF_1}
Given an $\wHyp{CFG}$ $G = (N,T,P,S,\mathcal{F})$, there can be constructed a $(\sigma,\rho,\pi)$-equivalent $\wHyp{CFG}$ $G' = (N',T,P',S,\mathcal{H})$ whose productions are of the forms $A \rightarrow \beta$, $A \rightarrow a$ or $A \rightarrow \epsilon$, $\beta \in N^+$, $A \in N$, $a \in T$, such that $L_{\sigma,\rho,\pi}(G) = L_{\sigma,\rho,\pi}(G')$, for every $(\sigma, \rho,\pi)$.
\end{lemma}

\begin{lemma}\label{Lem:wCSG_NF_1}
Given an $\wHyp{CSG}$ $G = (N,T,P,S,\mathcal{F})$, there can be constructed a $(\sigma,\rho,\pi)$-equivalent $\wHyp{CSG}$ $G' = (N',T,P',S,\mathcal{H})$ whose productions are of the forms $\alpha \rightarrow \beta$ or $A \rightarrow a$, $\alpha,\beta \in N^+$, $|\alpha| \leq |\beta|$, $A \in N$, $a \in T$, such that $L_{\sigma,\rho,\pi}(G) = L_{\sigma,\rho,\pi}(G')$, for every $(\sigma, \rho,\pi)$.
\end{lemma}

Using these normal forms, we can prove the construction of the $\$$-boundary form. For completeness, the following definition concerning $\$$-boundary is taken from Def. 4.5 of \cite{CG78b}.

\begin{definition}
An $\wHyp{PSG}$ ($\wHyp{CSG}$, resp.) with \emph{$\$$-boundary} is an $\omega$-grammar $G = (N \cup \{\$,S\},T,P$, $S_0,\mathcal{F})$, in which each production is of one of the following forms (1)-(4) ((1)-(3), resp.):
\begin{enumerate}
  \item $\alpha \rightarrow \beta$, $\alpha,\beta \in N^+$, (and $|\alpha| \leq |\beta|$ for $\wHyp{CSG}$)
  \item $S \rightarrow \$ \alpha$, $\alpha \in N^+$,
  \item $\$ A \rightarrow a \$$, $A \in N$, $a \in T$,
  \item $A \rightarrow \epsilon$, $A \in N$.\hfill$\Box$
\end{enumerate}
\end{definition}
The $\$$-boundary divides every sentential form into two parts. The left part consists of the generated string of terminals (never to be rewritten again). The right part consists of nonterminals to be rewritten.

The following lemma extends Thm. 4.6 of \cite{CG78b} where only 3-acceptance was considered (furthermore, the assumption in their proof was used without justification, while we provided a proof in Lemma \ref{Lem:wPSG_NF_1}).

\begin{lemma}\label{Lem:wPSG_NF_dol_bnd}
Given an $\wHyp{PSG}$ ($\wHyp{CSG}$, resp.), there can be constructed a $(\sigma,\rho,nl)$-equivalent $\wHyp{PSG}$ ($\wHyp{CSG}$, resp.) with $\$$-boundary, for every $(\sigma, \rho)$.
\end{lemma}
\proof Let $G = (N,T,P,S,\mathcal{F})$ be an $\wHyp{PSG}$. By Lemma \ref{Lem:wPSG_NF_1}, we assume $P$ = $P_1 \cup P_2 \cup P_3$, where $P_1 = \{p_k: \alpha \rightarrow \beta, \alpha,\beta \in N^+\}$, $P_2 = \{p_k: A \rightarrow a, A \in N, a \in T \}$, $P_3 = \{p_k: A \rightarrow \epsilon, A \in N \}$. There can be constructed a $(\sigma,\rho,nl)$-equivalent $\wHyp{PSG}$ $G' = (N \cup \{ \overline{a} ~|~ a \in T\} \cup \{S_1, \$\},T,P',S_1,\mathcal{H})$, where $P' = P_s \cup P_1 \cup P_2' \cup P_3 \cup P_4$, and $P_s = \{ S_1 \rightarrow \$ S \}$, $P_2' = \{p_k:  A \rightarrow \overline{a} ~|~p_k: A \rightarrow a \in P_2 \}$, $P_4 = \{ p_a: \$ \overline{a} \rightarrow a \$ ~|~ a \in T \}$.

Let $\mathcal{F} = \{F_i\}_{1\leq i \leq n}$, we construct the set $\mathcal{H}$ according to different acceptance modes:
\begin{enumerate}
\item $(\ran,\sqcap,nl)$-acceptance. $\mathcal{H} = \mathcal{F}$.
\item $(\ran,\subseteq,nl)$-acceptance. $\mathcal{H} = \{ F_i \cup P_s \cup P_4 \}_{1\leq i \leq n}$.
\item $(\ran,=,nl)$-acceptance. Let $\mathcal{H}_i = \{F_i \cup P_s \cup H ~|~ \emptyset \subset H \subseteq P_4 \}$, then $\mathcal{H} = \bigcup_{i = 1}^n \mathcal{H}_i$.
\item $(\inf,\sqcap,nl)$-acceptance. The same as (1).
\item $(\inf,\subseteq,nl)$-acceptance. $\mathcal{H} = \{ F_i \cup P_4 \}_{1\leq i \leq n}$.
\item $(\inf,=,nl)$-acceptance. Let $\mathcal{H}_i = \{F_i \cup H ~|~ \emptyset \subset H \subseteq P_4 \}$, then $\mathcal{H} = \bigcup_{i = 1}^n \mathcal{H}_i$.
\end{enumerate}
$G'$ generates the $\$$-boundary using $P_s$, then simulates $G$ using $P_1 \cup P_2' \cup P_3$ where each terminal $a$ of $G$ is replaced by nonterminal $\overline{a}$ of $G'$. Finally, $G'$ generates the $\omega$-word by moving $\$$ rightwards and replacing $\overline{a}$ by terminal $a$. It can be easily verified that $L_{\sigma,\rho,nl}(G) = L_{\sigma,\rho,nl}(G')$, for all $(\sigma, \rho, nl)$-acceptance modes.

If $G$ is an $\wHyp{CSG}$, $P_3$ above will be empty and $G'$ will be an $\wHyp{CSG}$ with $\$$-boundary. \hfill$\Box$

These important special forms discussed above can facilitate the proofs in the sequel.

\section{Leftmost Derivations of $\omega$-Grammar}
\label{Sec:wgrammar_lm}
In the case of leftmost derivation, we will show the equivalence of $\wHyp{RLG}$ and $\wHyp{FSA}$, and the equivalence of $\wHyp{CFG}$ and $\wHyp{PDA}$, as one may expect. Furthermore, for the leftmost derivation, the generative power of $\wHyp{CSG}$ or $\wHyp{PSG}$ is not greater than $\wHyp{PDA}$. In this section, most of the results are obtained by extending the results about the grammars on finite words.

\begin{theorem}\label{Thm:wRLG_lequ_wFSA}
$\mathcal{L}_{\sigma,\rho,l}(\wHyp{RLG}) = \mathcal{L}_{\sigma,\rho}(\wHyp{FSA})$, for every $(\sigma,\rho)$.
\end{theorem}
\proof This type of equation can be proved by showing the mutual inclusion of its two sides.

(i) $\mathcal{L}_{\sigma,\rho,l}(\wHyp{RLG}) \subseteq \mathcal{L}_{\sigma,\rho}(\wHyp{FSA})$. Let $G = (N, T, P, S_0, \mathcal{F})$ be an $\wHyp{RLG}$ with $N = \{S_k\}_{0 \leq k < |N|}$, $P = \{p_k\}_{1\leq k \leq |P|}$. Without loss of generality, we assume the productions are of the form $S_i \rightarrow a S_j$ or $S_i \rightarrow a$, $a \in T \cup \{\epsilon\}$ (by Lemma \ref{Lem:wRLG_no_Tstar_p}). Construct an $\wHyp{FSA}$ $A = (Q, T, \delta, q_0, \mathcal{H})$, where:
\begin{enumerate}
  \item $Q = \{q_0\} \cup \{ q_{k} ~|~ p_k: S_i \rightarrow \gamma \in P$, $\gamma \in (N \cup T)^* \}$, $q_0$ is the start state.
  \item $\delta(q_0, \epsilon) = \{ q_{k} ~|~ p_k: S_0 \rightarrow \gamma \in P \}$.
  \item $\delta(q_{k},a)$ contains $\{ q_{n} ~|~ p_n : S_j \rightarrow \gamma \in P \}$, for each production $p_k: S_i \rightarrow a S_j \in P$, where $a \in T \cup \{ \epsilon \}$. Note that $q_{k}$ reads one $a$, iff $p_k$ generates one $a$ by rewriting $S_i$.
\end{enumerate}
Obviously, in the bisimulation of $G$ and $A$, $G$ applies production $p_k$, iff $A$ uses a transition starting from $q_{k}$. Let $\mathcal{F} = \{F_m\}_{1\leq m \leq n}$, we construct the set $\mathcal{H}$ according to different acceptance modes:
\begin{enumerate}
\item $(\ran,\sqcap)$-acceptance. Let $H_m = \{q_k ~|~p_k \in F_m\}$, then $\mathcal{H} = \{ H_m\}_{1\leq m \leq n}$.
\item $(\ran,\subseteq)$-acceptance. Let $H_m = \{ q_0 \} \cup \{q_k ~|~p_k \in F_m\}$, then $\mathcal{H} = \{ H_m\}_{1\leq m \leq n}$.
\item $(\ran,=)$-acceptance. The same as (2).
\item $(\inf,\sqcap)$-acceptance. The same as (1).
\item $(\inf,\subseteq)$-acceptance. The same as (1).
\item $(\inf,=)$-acceptance. The same as (1).
\end{enumerate}
It can be easily verified that $L_{\sigma,\rho,l}(G) = L_{\sigma,\rho}(A)$.

(ii) $\mathcal{L}_{\sigma,\rho,l}(\wHyp{RLG}) \supseteq \mathcal{L}_{\sigma,\rho}(\wHyp{FSA})$. Let $A = (Q, \Sigma, \delta, q_0, \mathcal{F})$ be an $\wHyp{FSA}$ with $Q = \{q_k\}_{0 \leq k < |Q|}$. Construct an $\wHyp{RLG}$ $G = (N,\Sigma,P,S_0,\mathcal{H})$ where:
\begin{enumerate}
  \item for each $q_i \in Q$, there is a nonterminal $S_i \in N$, and $S_0$ is the start symbol.
  \item for each transition $\delta(q_i,a) \ni q_j$, where $a \in \Sigma \cup \{ \epsilon \}$, there is a production $p_{iaj}: S_i \rightarrow a S_j \in P$.
\end{enumerate}
We denote by $P_i$ the set of the production $p_{iaj} \in P$ for any $a \in \Sigma \cup \{ \epsilon \}$, $S_j \in N$. Obviously, in the bisimulation of $A$ and $G$, $A$ uses a transition starting from $q_{i}$, iff $G$ applies a production in $P_i$. Let $\mathcal{F} = \{F_k\}_{1\leq k \leq n}$, we construct the set $\mathcal{H}$ according to different acceptance modes:
\begin{enumerate}
\item $(\ran,\sqcap,l)$-acceptance. Let $H_k = \bigcup_{q_i \in F_k} P_i $, then $\mathcal{H} = \{ H_k\}_{1\leq k \leq n}$.
\item $(\ran,\subseteq,l)$-acceptance. The same as (1).
\item $(\ran,=,l)$-acceptance. Let $\mathcal{H}_k = \{ H \subseteq \bigcup_{q_i \in F_k} P_i ~|~\forall q_i \in F_k, H \cap P_i \neq \emptyset \}$, then $\mathcal{H} = \bigcup_{k=1}^n \mathcal{H}_k$.
\item $(\inf,\sqcap,l)$-acceptance. The same as (1).
\item $(\inf,\subseteq,l)$-acceptance. The same as (1).
\item $(\inf,=,l)$-acceptance. The same as (3).
\end{enumerate}
It can be easily verified that $L_{\sigma,\rho,l}(G) = L_{\sigma,\rho}(A)$.
\hfill$\Box$

\begin{theorem}\label{Thm:wCFG_lequ_wPDA}
$\mathcal{L}_{\sigma,\rho,l}(\wHyp{CFG}) = \mathcal{L}_{\sigma,\rho}(\wHyp{PDA})$, for every $(\sigma,\rho)$.
\end{theorem}
\proof (i) $\mathcal{L}_{\sigma,\rho,l}(\wHyp{CFG}) \subseteq \mathcal{L}_{\sigma,\rho}(\wHyp{PDA})$. Let $G = (N, T, P, S, \mathcal{F})$ be an $\wHyp{CFG}$ with $P = \{p_i\}_{1\leq i \leq |P|}$. Construct an $\wHyp{PDA}$ $D = (Q, T, \Gamma, \delta, q_0, S, \mathcal{H})$, where $\Gamma = N \cup T$, $Q = \{q_0\} \cup \{q_{i} ~|~p_i\in P \}$, $\delta$ is defined as follows:
\begin{enumerate}
  \item $\delta(q_0,a,a)=(q_0,\epsilon)$ for all $a \in T$,
  \item $\delta(q_0,\epsilon,A) \ni (q_{i},A)$ for $p_i: A \rightarrow \gamma \in P$,
  \item $\delta(q_{i},\epsilon,A)=(q_0,\gamma)$ for $p_i:A \rightarrow \gamma \in P$.
\end{enumerate}
Obviously, $D$ simulates the leftmost derivation of $G$. In the bisimulation of $G$ and $D$, $G$ applies production $p_i$, iff $D$ enters the state $q_i$. Let $\mathcal{F} = \{F_k\}_{1\leq k \leq n}$, we construct the set $\mathcal{H}$ according to different acceptance modes:
\begin{enumerate}
\item $(\ran,\sqcap)$-acceptance. Let $H_k = \{q_{i} ~|~p_i \in F_k\}$, then $\mathcal{H} = \{ H_k\}_{1\leq k \leq n}$.
\item $(\ran,\subseteq)$-acceptance. Let $H_k = \{q_0\} \cup \{q_{i} ~|~p_i \in F_k\}$, then $\mathcal{H} = \{ H_k\}_{1\leq k \leq n}$.
\item $(\ran,=)$-acceptance. The same as (2).
\item $(\inf,\sqcap)$-acceptance. The same as (1).
\item $(\inf,\subseteq)$-acceptance. The same as (2).
\item $(\inf,=)$-acceptance. The same as (2).
\end{enumerate}
It can be easily verified that $L_{\sigma,\rho,l}(G) = L_{\sigma,\rho}(D)$.

(ii) $\mathcal{L}_{\sigma,\rho,l}(\wHyp{CFG}) \supseteq \mathcal{L}_{\sigma,\rho}(\wHyp{PDA})$. Let $D = (Q, \Sigma, \Gamma, \delta, q_0, Z_0, \mathcal{F})$ be an $\wHyp{PDA}$. Construct an $\wHyp{CFG}$ $G = (N, \Sigma, P, S, \mathcal{H})$, where $N$ is the set of objects of the form $[q, B, r]$ (denoting popping $B$ from the stack by several transitions, switching the state from $q$ to $r$), $q, r \in Q$, $B \in \Gamma$, $P$ is the union of the following sets of productions:
\begin{enumerate}
  \item $P_s = \{S \rightarrow [q_0, Z_0, q_i] ~|~ q_i \in Q \}$.
  \item $P' = \{[q_i, B, q_j] \rightarrow a [q_{j_1}, B_1, q_{j_2}] [q_{j_2}, B_2, q_{j_3}] \cdots [q_{j_m}, B_m, q_{j}] ~|~\\ \delta (q_i, a, B) = (q_{j_1}, B_1B_2...B_m)$, $q_{j_2}$, ..., $q_{j_{m}}$, $q_j \in Q$, where $a \in \Sigma \cup \{\epsilon\}$, and $B, B_1, ..., B_m \in \Gamma \}$. (If $m=0$, then the production is $[q_i, B, q_{j_1}] \rightarrow a$.)
\end{enumerate}
We denote by $P_i$ the set of productions of the form $[q_i, B, q_j] \rightarrow \gamma$, for any $B \in \Gamma$, $q_j \in Q$, $\gamma \in N^* \cup \Sigma N^*$. Obviously, in the bisimulation of $D$ and $G$, $D$ uses a transition starting from $q_{i}$, iff $G$ applies a production in $P_i$.

Let $\mathcal{F} = \{F_k\}_{1\leq k \leq n}$, we construct the set $\mathcal{H}$ according to different acceptance modes:
\begin{enumerate}
\item $(\ran,\sqcap,l)$-acceptance. Let $H_k = \bigcup_{q_i \in F_k}P_{i}$, then $\mathcal{H} = \{ H_k\}_{1\leq k \leq n}$.
\item $(\ran,\subseteq,l)$-acceptance. Let $H_k = P_s \cup \bigcup_{q_i \in F_k}P_{i}$, then $\mathcal{H} = \{ H_k\}_{1\leq k \leq n}$.
\item $(\ran,=,l)$-acceptance. Let $\mathcal{H}_k = \{H \subseteq P ~|~ H \cap P_s \neq \emptyset$ and $\forall q_i \in F_k, H \cap P_{i} \neq \emptyset \}$, then $\mathcal{H} = \bigcup_{k = 1}^n \mathcal{H}_k$.
\item $(\inf,\sqcap,l)$-acceptance. The same as (1).
\item $(\inf,\subseteq,l)$-acceptance. The same as (1).
\item $(\inf,=,l)$-acceptance. Let $\mathcal{H}_k = \{H \subseteq P' ~|~ \forall q_i \in F_k, H \cap P_{i} \neq \emptyset \}$, then $\mathcal{H} = \bigcup_{k = 1}^n \mathcal{H}_k$.
\end{enumerate}
It can be easily verified that $L_{\sigma,\rho,l}(G) = L_{\sigma,\rho}(D)$. \hfill$\Box$

\begin{theorem}\label{Thm:wPSG_lequl_wCFG}
$\mathcal{L}_{\sigma,\rho,l}(\wHyp{PSG}) = \mathcal{L}_{\sigma,\rho,l}(\wHyp{CFG})$, for every $(\sigma,\rho)$.
\end{theorem}
\proof (i) $\mathcal{L}_{\sigma,\rho,l}(\wHyp{PSG}) \supseteq \mathcal{L}_{\sigma,\rho,l}(\wHyp{CFG})$ is trivial.

(ii) $\mathcal{L}_{\sigma,\rho,l}(\wHyp{PSG}) \subseteq \mathcal{L}_{\sigma,\rho,l}(\wHyp{CFG})$. We only need to prove $\mathcal{L}_{\sigma,\rho,l}(\wHyp{PSG}) \subseteq \mathcal{L}_{\sigma,\rho}(\wHyp{PDA})$ and the result follows from Thm. \ref{Thm:wCFG_lequ_wPDA}.

Let $G = (N, T, P, S, \mathcal{F})$ be an $\wHyp{PSG}$ with $P = \{p_i\}_{1\leq i \leq |P|}$. Construct an $\wHyp{PDA}$ $D = (Q, T, \Gamma, \delta, q_0', Z_0, \mathcal{H})$, where $\Gamma = N \cup T \cup \{Z_0\}$. Let $l$ be the maximal length of the left-hand sides of the productions of $P$, then $Q = \{q_0',q_0\} \cup \{q_{[i\alpha]} ~|~p_i\in P, \alpha \in \bigcup_{j=1}^l N^j \}$. $\delta$ is defined as follows:
\begin{enumerate}
  \item $\delta(q_0',\epsilon,Z_0)=(q_0,SZ_0)$,
  \item $\delta(q_0,a,a)=(q_0,\epsilon)$ for all $a \in T$,
  \item $\delta(q_0,\epsilon,A) \ni (q_{[iA]},\epsilon)$ if $p_i: A \gamma_1 \rightarrow \gamma \in P$, $A \in N$, $\gamma_1 \in N^*, \gamma \in (N \cup T)^*$,
  \item $\delta(q_{[i\alpha]},\epsilon,A)=(q_{[i\alpha A]},\epsilon)$ if $p_i: \alpha A \gamma_1 \rightarrow \gamma \in P$, $\alpha \in N^+$,
  \item $\delta(q_{[i\alpha]},\epsilon,X)=(q_0,\gamma X)$ if $p_i:\alpha \rightarrow \gamma \in P$, $X \in \Gamma$.
\end{enumerate}
In Item (3), when $D$ has the configuration $(q_0,A\beta)$, it guesses to apply $p_i:A \gamma_1 \rightarrow \gamma$. If the guess is wrong, $D$ blocks. Obviously, $D$ simulates the leftmost derivation of $G$. In the bisimulation, $G$ applies production $p_i: \alpha \rightarrow \gamma$, iff $D$ enters the state $q_{[i\alpha]}$.

We denote by $\pref(\alpha)$ the set of prefixes (length between 1 and $|\alpha|$) of finite word $\alpha$. Let $\mathcal{F} = \{F_k\}_{1\leq k \leq n}$, $Q_i = \{q_{[i\beta]} ~|~p_i:\alpha \rightarrow \gamma\in P$ and $\beta \in \pref(\alpha) \}$, we construct the set $\mathcal{H}$ according to different acceptance modes:
\begin{enumerate}
\item $(\ran,\sqcap)$-acceptance. Let $H_k = \{q_{[i\alpha]} ~|~p_i:\alpha \rightarrow \gamma \in F_k\}$, then $\mathcal{H} = \{ H_k\}_{1\leq k \leq n}$.
\item $(\ran,\subseteq)$-acceptance. Let $H_k = \{q_0',q_0\} \cup \bigcup_{p_i \in F_k} Q_i$, then $\mathcal{H} = \{ H_k\}_{1\leq k \leq n}$.
\item $(\ran,=)$-acceptance. The same as (2).
\item $(\inf,\sqcap)$-acceptance. The same as (1).
\item $(\inf,\subseteq)$-acceptance. Let $H_k = \{q_0\} \cup \bigcup_{p_i \in F_k} Q_i$, then $\mathcal{H} = \{ H_k\}_{1\leq k \leq n}$.
\item $(\inf,=)$-acceptance. The same as (5).
\end{enumerate}
It can be easily verified that $L_{\sigma,\rho,l}(G) = L_{\sigma,\rho}(D)$. \hfill$\Box$

\begin{theorem}\label{Thm:wCSG_lequl_wCFG}
$\mathcal{L}_{\sigma,\rho,l}(\wHyp{CSG}) = \mathcal{L}_{\sigma,\rho,l}(\wHyp{CFG})$, for $(\sigma,\rho) \not\in \{(\ran, \sqcap),(\ran, =)\}$.
\end{theorem}
\proof Note that the family of $\wHyp{CSG}$ includes $\epsilon$-production-free $\wHyp{CFG}$, and belongs to $\wHyp{PSG}$.\\
(i) $\mathcal{L}_{\sigma,\rho,l}(\wHyp{CSG}) \subseteq \mathcal{L}_{\sigma,\rho,l}(\wHyp{CFG})$ follows from $\mathcal{L}_{\sigma,\rho,l}(\wHyp{CSG}) \subseteq \mathcal{L}_{\sigma,\rho,l}(\wHyp{PSG})$ and Thm. \ref{Thm:wPSG_lequl_wCFG}.\\
(ii) $\mathcal{L}_{\sigma,\rho,l}(\wHyp{CFG}) \subseteq \mathcal{L}_{\sigma,\rho,l}(\wHyp{CSG})$ follows from Lemma \ref{Lem:wCFG_no_epsilon}. \hfill$\Box$

For the remaining two acceptance modes, we can only prove case (i) (the next theorem).

\begin{theorem}\label{Thm:wCSG_lequl_wCFG_1}
$\mathcal{L}_{\sigma,\rho,l}(\wHyp{CSG}) \subseteq \mathcal{L}_{\sigma,\rho,l}(\wHyp{CFG})$, for $(\sigma,\rho) \in \{(\ran, \sqcap),(\ran, =)\}$.
\end{theorem}
Unfortunately, whether the equivalence of $\wHyp{CSG}$ and $\wHyp{CFG}$ holds for the two acceptance modes is still an open problem. The difficulty comes from that, for the two modes, it is still unknown whether there can be constructed an $\epsilon$-production-free $\omega$-grammar for every $\wHyp{CFG}$ (as explained after Lemma \ref{Lem:wCFG_no_epsilon}).

Because of the equivalence of $(\sigma,\rho,l)$-accepting $\wHyp{CFG}$ and $(\sigma,\rho)$-accepting $\wHyp{PDA}$, we have the following corollary.
\begin{corollary}\label{Thm:wPSG_lequ_wPDA}
(i) $\mathcal{L}_{\sigma,\rho,l}(\wHyp{PSG}) = \mathcal{L}_{\sigma,\rho}(\wHyp{PDA})$, for every $(\sigma,\rho)$.

(ii) $\mathcal{L}_{\sigma,\rho,l}(\wHyp{CSG}) = \mathcal{L}_{\sigma,\rho}(\wHyp{PDA})$, for $(\sigma,\rho) \not\in \{(\ran, \sqcap),(\ran, =)\}$.

(iii) $\mathcal{L}_{\sigma,\rho,l}(\wHyp{CSG}) \subseteq \mathcal{L}_{\sigma,\rho}(\wHyp{PDA})$, for $(\sigma,\rho) \in \{(\ran, \sqcap),(\ran, =)\}$.
\end{corollary}

\section{Normal Derivations of $\omega$-Grammar}
\label{Sec:wgrammar_nl}
In the case of normal derivation, we will show the equivalence of $\wHyp{RLG}$ and $\wHyp{FSA}$, and the equivalence of $\wHyp{PSG}$ and $\wHyp{TM}$, as one may expect. Furthermore, for normal derivation, the generative power of $\wHyp{CFG}$ is not greater than (may be strictly included in, or equal to) $\wHyp{PDA}$. The generative power of $\wHyp{CSG}$ is also equal to $\wHyp{TM}$. In this section, most of the results are obtained by extending the results about the grammars on finite words.

\begin{theorem}\label{Thm:wRLG_equ_wFSA}
$\mathcal{L}_{\sigma,\rho,nl}(\wHyp{RLG}) = \mathcal{L}_{\sigma,\rho,l}(\wHyp{RLG}) = \mathcal{L}_{\sigma,\rho}(\wHyp{FSA})$, for every $(\sigma,\rho)$.
\end{theorem}
\proof It is trivial, since every derivation is a leftmost derivation for $\wHyp{RLG}$. The second equation is taken from Thm. \ref{Thm:wRLG_lequ_wFSA}. \hfill$\Box$

To study the generative power of $\wHyp{CFG}$, we must have in mind the following fact. Recall that for $CFG$ on finite words, given $G \in CFG$, the language generated by leftmost derivations and the one generated by normal derivations are equal, i.e., $L_l(G) = L_{nl}(G)$. For $\wHyp{CFG}$, given $G \in \wHyp{CFG}$, if we do not take into account the production repetition sets $\mathcal{F}$, then $G$ generates leftmostly $u \in T^\omega$, iff $G$ generates $u$ in a normal derivation, because $u$ is the leftmost substring consisting of terminals in a sentential form. It is important to note that, in a normal derivation of $\wHyp{CFG}$, there may exist some substrings of terminals that are obtained by rewriting some nonterminals in the unreached part of the sentential form (does not contribute to $u$), then its possible impact on the derivation lies only in its set of applied productions. We formally define the reached part and the unreached part of a sentential form as follows.

\begin{definition}
Let $G = (N,T,P,S,\mathcal{F})$ be an $\wHyp{CFG}$ and $V = N \cup T$. Let $d$ be an infinite derivation in $G$, $d: \alpha_1 \Rightarrow \alpha_2 \Rightarrow \cdots \Rightarrow \alpha_i \Rightarrow \cdots$. Every sentential form $\alpha_i$ can be decomposed into $\alpha_i = \beta_i\gamma_i$, and the derivation starting from $\alpha_i$ can be decomposed into $d_{\beta_i}: \beta_i \Rightarrow \beta_{i+1} \Rightarrow \cdots$ and $d_{\gamma_i}: \gamma_i \Rightarrow \gamma_{i+1} \Rightarrow \cdots$ where $\alpha_k = \beta_k \gamma_k$ for $k \geq i$, such that
\begin{enumerate}
  \item for every nonterminal $A$ in $\beta_i$ s.t. $\beta_i = \gamma A \gamma'$, $\gamma, \gamma' \in V^*$, $d_{\beta_i}$ rewrites $\gamma \Rightarrow^* T^*$,
  \item for every $k \geq i$, $\beta_k \in V^*NV^*$.
\end{enumerate}
We say $\beta_i$ and $\gamma_i$ are the \emph{reached part} and the \emph{unreached part} of $\alpha_i$, respectively. \hfill$\Box$
\end{definition}
In words, all the nonterminals in the reached part $\beta_i$ will be rewritten in the derivation, but $\beta_i$ will never be completely rewritten to be a string of terminals, thus $\gamma_i$ will not be reached by the generated $\omega$-word.

If a string $\gamma$ appears in the unreached part of a sentential form, then it does not contribute to the terminals in the generated $\omega$-word, but only contributes to the set of productions used (by its transient sets) and the set of productions that appear infinitely often (by its self-providing sets).

\begin{definition}
Let $G = (N,T,P,S,\mathcal{F})$ be an $\wHyp{CFG}$. For any $\gamma \in (N \cup T)^*$, the class of \emph{self-providing sets} $SP(\gamma)$ and the class of \emph{transient sets} $TR(\gamma)$ are defined as
\begin{align*}
SP(\gamma) &= \{ D \subseteq P ~|~ \text{there exists an infinite nl-derivation $d$ }\\
           &~~~~~~~~~~~~~~~~~~\text{starting in $\gamma$ s.t. $\inf(d_P) = D$} \} \\
TR(\gamma) &= \{ D \subseteq P ~|~ \text{there exists a finite nl-derivation $d: \gamma \Rightarrow^* \gamma'$ }\\
           &~~~~~~~~~~~~~~~~~~\text{for some $\gamma' \in (N \cup T)^*$ s.t. $\ran(d_P) = D$} \}
\end{align*}\hfill$\Box$
\end{definition}
It follows immediately that
\begin{align*}
SP(\alpha\beta) &= SP(\alpha) \cup SP(\beta) \cup \{H_1 \cup H_2 ~|~H_1 \in SP(\alpha), H_2 \in SP(\beta) \}\\
TR(\alpha\beta) &= TR(\alpha) \cup TR(\beta) \cup \{H_1 \cup H_2 ~|~H_1 \in TR(\alpha), H_2 \in TR(\beta) \}
\end{align*}
Using the above concepts, we are ready to show the nl-derivation of $\wHyp{CFG}$ can be simulated by the computation of $\wHyp{PDA}$.

\begin{theorem}\label{Thm:wCFG_subseteql_wCFG}
$\mathcal{L}_{\sigma,\rho,nl}(\wHyp{CFG}) \subseteq \mathcal{L}_{\sigma,\rho}(\wHyp{PDA})$, for every $(\sigma,\rho)$.
\end{theorem}
\proof Given a $(\sigma,\rho,nl)$-accepting $\wHyp{CFG}$ $G = (N,T,P,S,\mathcal{F})$, we only need to show how to construct an $\wHyp{PDA}$ $D = (Q,T,\Gamma,\delta,q,S,\mathcal{H})$, such that $G$ and $D$ accept exactly the same $\omega$-language.

Without loss of generality (because of the closure property under union), we may assume $\mathcal{F}$ consists of only one repetition set, denoted by $F$. We may also assume that $P = P_1 \cup P_2$, where $P_1$ are of the form $A \rightarrow \beta$, $\beta \in N^+$, and $P_2$ are of the form $A \rightarrow a$, $a \in \Sigma \cup \{ \epsilon \}$ (by Lemma \ref{Lem:wCFG_NF_1}). Let $TR_1(\gamma) = \{ D \subseteq F ~|~$ there exists a finite nl-derivation $d: \gamma \Rightarrow^* \gamma'$ for some $\gamma' \in (N \cup T)^*$ s.t. $\ran(d_P) = D \}$. Note that $TR_1(\gamma)$ is a set of subsets of $F$. It is different from $TR(\gamma)$ since it only includes the sets whose elements are all in $F$, i.e. $\{ D \subseteq F ~|~D \in TR(\gamma)\}$.

(Case 1) $(\inf,\subseteq,nl)$-acceptance. Assume $P =\{p_i\}_{1 \leq i \leq |P|}$, we construct an $\wHyp{PDA}$ $D$ with $Q = \{q \} \cup \{q_i ~|~ p_i \in P\} $, $\Gamma = N \cup T$, $\delta$ is defined as follows:
\begin{enumerate}
  \item $\delta (q,\epsilon,A) \ni (q_i,A)$ for $p_i: A \rightarrow \gamma \in P$,
  \item $\delta (q_i,\epsilon,A) = (q,\gamma)$ for $p_i: A \rightarrow \gamma \in P$,
  \item $\delta (q,a,a) = (q,\epsilon)$ for all $a \in T$.
\end{enumerate}
Let $\mathcal{H} = \{\{ q \} \cup \{q_i ~|~ p_i \in F \}\}$. Obviously, $D$ simulates the derivation of $G$ in the reached part. Note that $D$ may simulate less productions than $G$, due to the derivation in the unreached part. But this fact does not affect the $(\inf,\subseteq,nl)$-accepted $\omega$-languages. It can be easily verified that $L_{\inf,\subseteq,nl}(G) = L_{\inf,\subseteq}(D)$.

(Case 2) $(\ran,\subseteq,nl)$-acceptance. The same as Case 1.

(Case 3) $(\inf,\sqcap,nl)$-acceptance. We construct an $\wHyp{PDA}$ $D$ with $Q = \{q_0\} \times \{0,1,2\}$, $\Gamma = N \cup \{Z\}$ where $Z \not\in N $, and $q = [q_0, 0]$. The second component of $Q$ is used to remember whether some productions in $F$ appear once again (value 1) or infinitely often (value 2). If it equals 1, it returns to 0 in the next state, whereas the value 2 will hold forever. Thus we only need to guarantee that, in a legal run, the value 1 or 2 appears infinitely often. Define the function $f$ as follows: if $p \in F$, then $f(0,p) = 1$, else $f(0,p) = 0$; for $p \in P$, $f(1,p) = 0$ and $f(2,p) = 2$. The transition function $\delta$ is defined as follows.
\begin{enumerate}
  \item $\delta ([q_0,i],a,A) = ([q_0,f(i,p)],\epsilon)$, if $p: A \rightarrow a \in P_2$, $a \in \Sigma \cup \{\epsilon\}$,
  \item $\delta ([q_0,i],\epsilon,A) \ni ([q_0,f(i,p)],\beta)$, if $p: A \rightarrow \beta \in P_1$, $\beta \in N^+$,
  \item $\delta ([q_0,i],\epsilon,A) \ni ([q_0,2],\gamma_1 Z)$, if $A \rightarrow \gamma_1\gamma_2 \in P_1$, $\gamma_1,\gamma_2 \neq \epsilon$, and $\exists K \in SP(\gamma_2)$ s.t. $F \cap K \neq \emptyset$, (nondeterministically choose $\gamma_2$ as an unreached part that applies some productions in $F$ infinitely often.)
  \item $\delta ([q_0,i],\epsilon,A) \ni ([q_0,1],\gamma_1 Z)$, if $A \rightarrow \gamma_1\gamma_2 \in P_1$, $\gamma_1,\gamma_2 \neq \epsilon$, and $\exists K \in TR(\gamma_2)$ s.t. $F \cap K \neq \emptyset$.
\end{enumerate}
Let $\mathcal{H} = \{\{ [q_0,1],[q_0,2] \}\}$. The infinite appearances of $[q_0,1]$ mean some productions in $F$ appear infinitely often in the derivation of $G$, while $[q_0,2]$ means some productions in $F$ appear infinitely often in a certain unreached part of the derivation. It can be easily verified that $L_{\inf,\sqcap,nl}(G) = L_{\inf,\sqcap}(D)$.

(Case 4) $(\ran,\sqcap,nl)$-acceptance. We construct an $\wHyp{PDA}$ $D$ with $Q = \{q_0\} \times \{0,1\}$, $\Gamma = N \cup \{Z\}$ where $Z \not\in N $, and $q = [q_0, 0]$. Whenever a production in $F$ is applied, $D$ enters $[q_0,1]$, and the value 1 will hold forever. Define the function $f$ as follows: if $p \in F$, then $f(0,p) = 1$, else $f(0,p) = 0$; for $p \in P$, $f(1,p) = 1$. The transition function $\delta$ is defined as follows.
\begin{enumerate}
  \item $\delta ([q_0,i],a,A) \ni ([q_0,f(i,p)],\epsilon)$, if $p: A \rightarrow a \in P_2$, $a \in \Sigma \cup \{\epsilon\}$,
  \item $\delta ([q_0,i],\epsilon,A) \ni ([q_0,f(i,p)],\beta)$, if $p: A \rightarrow \beta \in P_1$, $\beta \in N^+$,
  \item $\delta ([q_0,i],\epsilon,A) \ni ([q_0,1],\gamma_1 Z)$, if $A \rightarrow \gamma_1\gamma_2 \in P_1$, $\gamma_1,\gamma_2 \neq \epsilon$, and $\exists K \in TR(\gamma_2)$ s.t. $F \cap K \neq \emptyset$, (nondeterministically choose $\gamma_2$ as an unreached part that applies some productions in $F$.)
\end{enumerate}
Let $\mathcal{H} = \{\{ [q_0,1] \}\}$. Obviously, in the bisimulation of $D$ and $G$, $D$ enters $[q_0,1]$, iff $G$ applies a production in $F$. It can be easily verified that $L_{\ran,\sqcap,nl}(G) = L_{\ran,\sqcap}(D)$.

(Case 5) $(\inf,=,nl)$-acceptance. We construct an $\wHyp{PDA}$ $D$ with $Q = \{q_0\} \times 2^{2^F} \cup \{q_1\} \times 2^F \times 2^F \cup \{\overline{q}\} \times 2^F$, $\Gamma = N \cup \{Z\}$ where $Z \not\in N $, and $q = [q_0, \emptyset]$. The second component of $Q$ is used to remember the set of productions that have appeared infinitely often in the unreached part of a derivation. The transition function $\delta$ is defined as follows.
\begin{enumerate}
  \item $\delta ([q_0,H],a,A) \ni ([q_0,H],\epsilon)$ for $H \subseteq 2^F$, if $A \rightarrow a \in P_2$,
  \item $\delta ([q_0,H],\epsilon,A) \ni ([q_0,H],\beta)$ for $H \subseteq 2^F$, if $A \rightarrow \beta \in P_1$,
  \item $\delta ([q_0,H],\epsilon,A) \ni ([q_0,H_1],\gamma_1 Z)$ for $H \subseteq 2^F$, if $A \rightarrow \gamma_1\gamma_2 \in P_1$, $\gamma_1,\gamma_2 \neq \epsilon$, and $H_1 = \{K_1 \cup K_2 ~|~ K_1 \in H, K_2 \in SP(\gamma_2) \}$, (nondeterministically choose $\gamma_2$ as an unreached part, and accumulate the productions that appear infinitely often in rewriting $\gamma_2$.)
  \item $\delta ([q_0,H],\epsilon,A) \ni ([q_1,F-K,\emptyset],A)$ for $H \subseteq 2^F$, $A \in N$, and $K \in H$, (start the derivation of the reached part using only productions that appear infinitely often. The third component of the state is used to accumulate productions applied infinitely often hereafter. $F-K$ computes the productions needed to appear infinitely often hereafter, since the productions in $K$ have appeared infinitely often in the unreached part.)
  \item $\delta ([q_1,K,H],a,A) \ni ([q_1,K,H \cup \{p\}],\epsilon)$ for $K,H \subseteq F$, if $p: A \rightarrow a \in P_2$ and $p \in F$, (accumulate a production that appears once.)
  \item $\delta ([q_1,K,H],\epsilon,A) \ni ([q_1,K,H \cup \{p\}],\beta)$ for $K,H \subseteq F$, if $p: A \rightarrow \beta \in P_1$ and $p \in F$, (accumulate a production that appears once.)
  \item $\delta ([q_1,K,H],\epsilon,A) \ni ([q_1,K,H \cup H_1 \cup \{p\}],\gamma_1 Z)$ for $K,H \subseteq F$, $H_1 \in TR_1(\gamma_2)$, if $p: A \rightarrow \gamma_1\gamma_2 \in P_1$, $\gamma_1,\gamma_2 \neq \epsilon$, and $p \in F$, (accumulate the productions that appear in the unreached part $\gamma_2$.)
  \item $\delta ([q_1,K,H],\epsilon,A) \ni ([\overline{q},K],A)$ for $K,H \subseteq F$, $A \in N$, if $K \subseteq H$, (when all the remaining productions that are required to appear infinitely often have been accumulated in $H$, $D$ enters $\overline{q}$.)
  \item $\delta ([\overline{q},K],\epsilon,A) \ni ([q_1,K,\emptyset],A)$ for $K \subseteq F$, $A \in N$. (restart accumulating, because each production in $K$ has been used at least once since the last time $D$ had entered $[q_1,K,\emptyset]$.)
\end{enumerate}
Let $\mathcal{H} = \{\{ \overline{q} \times 2^F \}\}$, it can be easily verified that, there can be constructed an $\wHyp{PDA}$ $D'$ from $D$ by only modifying $\mathcal{H}$, such that $L_{\inf,=,nl}(G) = L_{\inf,\sqcap}(D) = L_{\inf,=}(D')$.

(Case 6) $(\ran,=,nl)$-acceptance. We construct an $\wHyp{PDA}$ $D$ with $Q = \{q_0\} \times 2^F$, $\Gamma = N \cup \{Z\}$ where $Z \not\in N $, and $q = [q_0, \emptyset]$. The transition function $\delta$ is defined as follows.
\begin{enumerate}
  \item $\delta ([q_0,H],a,A) \ni ([q_0,H \cup \{p\}],\epsilon)$ for $H \subseteq F$, if $p: A \rightarrow a \in P_2$ and $p \in F$,
  \item $\delta ([q_0,H],\epsilon,A) \ni ([q_0,H \cup \{p\}],\beta)$ for $H \subseteq F$, if $p: A \rightarrow \beta \in P_1$ and $p \in F$,
  \item $\delta ([q_0,H],\epsilon,A) \ni ([q_0,H_1],\gamma_1 Z)$ for $H \subseteq F$, if $p: A \rightarrow \gamma_1\gamma_2 \in P_1$ and $p \in F$, $\gamma_1,\gamma_2 \neq \epsilon$, and $H_1 \in \{H \cup \{p\} \cup K ~|~ K \in TR_1(\gamma_2) \}$. (nondeterministically choose $\gamma_2$ as an unreached part.)
\end{enumerate}
Let $\mathcal{H} = \{ H \subseteq Q ~|~ [q_0,F] \in H\}$. Note that if $G$ applies a production outside $F$, $D$ blocks in the simulation. It can be easily verified that $L_{\ran,=,nl}(G) = L_{\ran,=}(D)$.
\hfill$\Box$

The above result also means $\mathcal{L}_{\sigma,\rho,nl}(\wHyp{CFG}) \subseteq \mathcal{L}_{\sigma,\rho,l}(\wHyp{CFG})$ by Thm. \ref{Thm:wCFG_lequ_wPDA}. Now we consider whether the proper inclusion or the equivalence holds.

\begin{theorem}\label{Thm:wCFG_subset_eq_l_wCFG}
(i) $\mathcal{L}_{\sigma,\rho,nl}(\wHyp{CFG}) \subset \mathcal{L}_{\sigma,\rho,l}(\wHyp{CFG})$, for $(\sigma,\rho) \in \{ (\inf,\sqcap), (\inf,=)\}$.\\
(ii) $\mathcal{L}_{\sigma,\rho,nl}(\wHyp{CFG}) = \mathcal{L}_{\sigma,\rho,l}(\wHyp{CFG})$, for $(\sigma,\rho) \in \{ (\ran,\subseteq), (\inf,\subseteq)\}$.
\end{theorem}
\proof We consider various acceptance modes one by one.
\begin{enumerate}
  \item $\mathcal{L}_{\inf,=,nl}(\wHyp{CFG}) \subset \mathcal{L}_{\inf,=,l}(\wHyp{CFG})$. It was proved that there exists an $\omega$-language $L = \{a^nb^n ~|~ n \geq 1\}^\omega$, such that $L \in \mathcal{L}_{\inf,=,l}(\wHyp{CFG})$, but $L \not\in \mathcal{L}_{\inf,=,nl}(\wHyp{CFG})$ (see Proposition 4.3.6 of \cite{CG77b}). It follows that $\mathcal{L}_{\inf,=,l}(\wHyp{CFG}) \nsubseteq \mathcal{L}_{\inf,=,nl}(\wHyp{CFG})$.
  \item $\mathcal{L}_{\inf,\sqcap,nl}(\wHyp{CFG}) \subset \mathcal{L}_{\inf,\sqcap,l}(\wHyp{CFG})$. Consider again $L = \{a^nb^n ~|~ n \geq 1\}^\omega$. On the one hand, we have $L \in \mathcal{L}_{\inf,\sqcap,l}(\wHyp{CFG})$, since $\mathcal{L}_{\inf,=,l}(\wHyp{CFG}) = \mathcal{L}_{\inf,\sqcap,l}(\wHyp{CFG})$ by Thm. \ref{Thm:wCFG_lequ_wPDA} and Thm. \ref{Thm:Gen_X_wAutomata}. On the other hand, $L \not\in \mathcal{L}_{\inf,\sqcap,nl}(\wHyp{CFG})$, because it is easy to prove $\mathcal{L}_{\inf,\sqcap,nl}(\wHyp{CFG}) \subseteq \mathcal{L}_{\inf,=,nl}(\wHyp{CFG})$. Therefore, it follows that\\ $\mathcal{L}_{\inf,\sqcap,l}(\wHyp{CFG}) \nsubseteq \mathcal{L}_{\inf,\sqcap,nl}(\wHyp{CFG})$.
  \item $\mathcal{L}_{\sigma,\subseteq,nl}(\wHyp{CFG}) = \mathcal{L}_{\sigma,\subseteq,l}(\wHyp{CFG})$, for $\sigma \in \{\ran,\inf\}$. Note that for any $G \in \wHyp{CFG}$, $L_{\sigma,\subseteq,nl}(G) = L_{\sigma,\subseteq,l}(G)$, thanks to the subtle semantics of the relation $\subseteq$. Therefore, it is easy to show $\mathcal{L}_{\sigma,\subseteq,nl}(\wHyp{CFG}) = \mathcal{L}_{\sigma,\subseteq,l}(\wHyp{CFG})$. \hfill$\Box$
\end{enumerate}

Unfortunately, for $(\sigma,\rho) \in \{(\ran, \sqcap),(\ran, =)\}$, whether proper inclusion or equivalence holds is still an open problem. The difficulty comes from two folds. First, we lack some examples like $L$ to prove the proper inclusion. Second, it is not easy to establish the relationships with other acceptance modes. Thus we cannot infer the result from the known results.

Now we consider the generative power of $\wHyp{CSG}$ and $\wHyp{PSG}$ together by proving the following lemmas.

\begin{lemma}\label{Thm:wTM_subset_wCSG}
$\mathcal{L}_{\sigma,\rho}(\wHyp{TM}) \subseteq \mathcal{L}_{\sigma,\rho,nl}(\wHyp{CSG})$, for every $(\sigma,\rho)$.
\end{lemma}
\proof Let $M = (Q, \Sigma, \Gamma, \delta, q_0, \mathcal{F})$ be an $\wHyp{TM}$. Construct an $\wHyp{CSG}$ $G = (N, \Sigma, P, S, \mathcal{H})$, where $N = \Sigma \times \Gamma \cup Q \cup \{ \$, S, S_1\}$, $P$ contains the following productions:
\begin{enumerate}
  \item $S \rightarrow \$ q_0 S_1$,
  \item $S_1 \rightarrow [a,a] S_1$, for every $a \in \Sigma$,
  \item $q[a,A] \rightarrow [a,C]p$, if $\delta(q,A) \ni (p,C,R)$ for every $a \in \Sigma$,
  \item $[b,B]q[a,A] \rightarrow p[b,B][a,C]$, if $\delta(q,A) \ni (p,C,L)$ for every $a,b \in \Sigma$, $B \in \Gamma$,
  \item $q[a,A] \rightarrow p[a,C]$, if $\delta(q,A) \ni (p,C,S)$ for every $a \in \Sigma$,
  \item $\$[a,A] \rightarrow a\$$, for every $a \in \Sigma$, $A \in \Gamma$.
\end{enumerate}
We denote by $P_i$ the set of productions of type (i) above. For every $q \in Q$, we denote by $P_{q}$ the set of productions in which $q$ appears on the left-hand side.

Productions $P_2$ can generate the input $\omega$-word. The first component of $\Sigma \times \Gamma$ is used to record the input symbol, and the second is used to simulate $M$. $M$ has a c.n.o. run on an infinite $\omega$-word, iff $G$ can generate the $\omega$-word by using some productions in $P_6$ infinitely often.

Let $\mathcal{F} = \{F_k\}_{1\leq k \leq n}$, we construct the set $\mathcal{H}$ according to different acceptance modes:
\begin{enumerate}
\item $(\ran,\sqcap,nl)$-acceptance. Let $H_k = \bigcup_{q \in F_k}P_{q}$, then $\mathcal{H} = \{ H_k\}_{1\leq k \leq n}$.
\item $(\ran,\subseteq,nl)$-acceptance. Let $H_k = P_1 \cup P_2 \cup P_6 \cup \bigcup_{q \in F_k}P_{q}$, then $\mathcal{H} = \{ H_k\}_{1\leq k \leq n}$.
\item $(\ran,=,nl)$-acceptance. Let $\mathcal{H}_k = \{H \subseteq P_1 \cup P_2 \cup P_6 \cup \bigcup_{q \in F_k} P_q ~|~ P_1 \subseteq H$ and $H \cap P_2 \neq \emptyset$ and $H \cap P_6 \neq \emptyset$ and $\forall q \in F_k, H \cap P_{q} \neq \emptyset \}$, then $\mathcal{H} = \bigcup_{k = 1}^n \mathcal{H}_k$.
\item $(\inf,\sqcap,nl)$-acceptance. The same as (1).
\item $(\inf,\subseteq,nl)$-acceptance. Let $H_k = P_2 \cup P_6 \cup \bigcup_{q \in F_k}P_{q}$, then $\mathcal{H} = \{ H_k\}_{1\leq k \leq n}$.
\item $(\inf,=,nl)$-acceptance. Let $\mathcal{H}_k = \{H \subseteq P_2 \cup P_6 \cup \bigcup_{q \in F_k} P_q ~|~ H \cap P_2 \neq \emptyset$ and $H \cap P_6 \neq \emptyset$ and $\forall q \in F_k, H \cap P_{q} \neq \emptyset \}$, then $\mathcal{H} = \bigcup_{k = 1}^n \mathcal{H}_k$.
\end{enumerate}
It can be easily verified that $L_{\sigma,\rho,nl}(G) = L_{\sigma,\rho}(M)$. \hfill$\Box$

\begin{lemma}\label{Thm:wPSG_subset_2wTM}
$\mathcal{L}_{\sigma,\rho,nl}(\wHyp{PSG}) \subseteq \mathcal{L}_{\sigma,\rho}(\HypwHyp{2}{TM})$, for every $(\sigma,\rho)$.
\end{lemma}
\proof Let $G = (N, T, P, S, \mathcal{F})$ be an $\wHyp{PSG}$. By Lemma \ref{Lem:wPSG_NF_dol_bnd}, we may assume that $G$ is an $\wHyp{PSG}$ with $\$$-boundary. Construct a $\HypwHyp{2}{TM}$ $M = (Q, T, \Gamma, \delta, q_0, \mathcal{H})$ where $Q = Q' \cup \{q_0, q_1, q_D\} \cup \{q_p: p \in P\}$, $Q'$ is a set of working states and $q_D$ is a dead state (no further transitions). The machine has two tapes. The first tape contains the input $u \in T^\omega$, while on the second tape $M$ simulates nondeterministically a derivation in $G$. $M$ starts with writing $S$ in the first square of the second tape. For every production $p$ in $P$ there is a corresponding state $q_p$ in $Q$, entered by $M$ every time production $p$ is simulated on the second tape. If $M$ cannot find a production to simulate, then $M$ enters the dead state $q_D$. Furthermore, each time $M$ simulates a production of the form $\$A \rightarrow a\$$, the terminal $ a\in \Sigma$ is checked against the letter pointed to on the first tape. If there is a match, $M$ enters state $q_1$, moves both the two reading heads of the two tapes one square to the right and then proceeds with the simulation. Otherwise, $M$ enters the dead state $q_D$. Note that the reading head on the first tape of $M$ moves one square to the right, iff $G$ adds one terminal to the generated $\omega$-word.

Let $\mathcal{F} = \{F_i\}_{1\leq i \leq n}$, we construct the set $\mathcal{H}$ according to different acceptance modes:
\begin{enumerate}
\item $(\ran,\sqcap)$-acceptance. Let $H_i = \{q_p ~|~ p \in F_i \} $, then $\mathcal{H} = \{ H_i \}_{1\leq i \leq n}$.
\item $(\ran,\subseteq)$-acceptance. Let $H_i = Q' \cup \{q_0,q_1\} \cup \{q_p ~|~ p \in F_i \} $, then $\mathcal{H} = \{ H_i \}_{1\leq i \leq n}$.
\item $(\ran,=)$-acceptance. Let $\mathcal{H}_i = \{ H \cup \{q_0,q_1\} \cup \{q_p ~|~ p \in F_i \} ~|~ H \subseteq Q' \}$, then $\mathcal{H} = \bigcup_{i = 1}^n \mathcal{H}_i$.
\item $(\inf,\sqcap)$-acceptance. The same as (1).
\item $(\inf,\subseteq)$-acceptance. Let $H_i = Q' \cup \{q_1\} \cup \{q_p ~|~ p \in F_i \} $, then $\mathcal{H} = \{ H_i \}_{1\leq i \leq n}$.
\item $(\inf,=)$-acceptance. Let $\mathcal{H}_i = \{ H \cup \{q_1\} \cup \{q_p ~|~ p \in F_i \} ~|~ H \subseteq Q' \}$, then $\mathcal{H} = \bigcup_{i = 1}^n \mathcal{H}_i$.
\end{enumerate}
It can be easily verified that $L_{\sigma,\rho,nl}(G) = L_{\sigma,\rho}(M)$. \hfill$\Box$

Note that the proof above has two important differences from the proof of Thm. 5.1 in \cite{CG78b} in which only the 3-accepting (i.e., $(\inf,=,nl)$-accepting) $\omega$-grammar was considered. The first difference is that we use two tapes rather than two tracks. Because if we used two tracks, except the $(\inf,=)$-acceptance case, for any input $u \in T^\omega$, the $\wHyp{TM}$ $M$ would have a c.n.o. run on $u$ satisfying the acceptance condition by applying an infinite computation as follows: $M$ generates only finite (maybe zero) terminals on the leftmost side of the second track, and then never uses the productions $\$ A \rightarrow a \$$ any more. Instead, $M$ may always rewrite the rightmost nonterminals on the second track, leading to a c.n.o. run. Thus, $M$ would accept $L_{\sigma,\rho}(M) = T^\omega$. The second difference is that we use $q_D$ instead of the traverse state $q_T$, because the latter brings inconvenience for $(\ran,\sqcap)$-acceptance: for any input $u \in T^\omega$, $M$ may simulate certain $p \in F_i$ once, then enter $q_T$ by a dismatch when comparing the terminals on the two tapes, and the run is a c.n.o. run and thus accepted. Therefore, $M$ would accept $L_{\ran,\sqcap}(M) = T^\omega$. In order to provide a uniform constructive proof for all the acceptance modes, we give our new proof by modifying the proof of Thm. 5.1 in \cite{CG78b}.

By the above lemmas and Thm. \ref{Thm:mwTM_equ_wTM}, we showed the fact $\mathcal{L}_{\sigma,\rho}(\wHyp{TM}) \subseteq$ $\mathcal{L}_{\sigma,\rho,nl}(\wHyp{CSG}) \subseteq$ $\mathcal{L}_{\sigma,\rho,nl}(\wHyp{PSG}) \subseteq$ $\mathcal{L}_{\sigma,\rho}(\HypwHyp{2}{TM}) \subseteq $ $\mathcal{L}_{\sigma,\rho}(\mwHyp{TM}) = \mathcal{L}_{\sigma,\rho}(\wHyp{TM})$. Thus all the elements in the formula are equivalent. We have the following theorems.

\begin{theorem}\label{Thm:wPSG_equ_wTM}
$\mathcal{L}_{\sigma,\rho,nl}(\wHyp{PSG}) = \mathcal{L}_{\sigma,\rho}(\wHyp{TM})$, for every $(\sigma,\rho)$.
\end{theorem}

\begin{theorem}\label{Thm:wCSG_equ_wTM}
$\mathcal{L}_{\sigma,\rho,nl}(\wHyp{CSG}) = \mathcal{L}_{\sigma,\rho}(\wHyp{TM})$, for every $(\sigma,\rho)$.
\end{theorem}

\section{Related Work}
\label{Sec:rwork}
Cohen only focused on the five types of $i$-accepting $\omega$-automaton and the 3-accepting $\omega$-grammar \cite{CG77a}, and mainly discussed $\wHypV{CFG}$ with variable repetition sets rather than the one with production repetition sets \cite{CG77b}. Therefore, in our notation, Cohen actually studied some relationships between $(\inf,=,\pi)$-accepting $\omega$-grammars and $(\inf,=)$-accepting $\omega$-automata (i.e., 3-acceptance), including $(\inf,=,\pi)$-accepting $\wHypV{CFG}$ and $(\inf,=)$-accepting $\wHyp{PDA}$ \cite{CG77a,CG77b}, $(\inf,=,nl)$-accepting $\wHyp{PSG}$ and $(\inf,=)$-accepting $\wHyp{TM}$ \cite{CG78b}. Furthermore, Cohen studied also some special forms and normal forms of some $(\inf,=,\pi)$-accepting $\omega$-grammars.

We extended Cohen's work in the following aspects. First, we gave clean and uniform definitions for $\omega$-automata, $\omega$-grammars, and the $\omega$-languages accepted by various acceptance modes. These notations help us to understand the results more clearly. Second, we examined the $\omega$-grammars beyond $(\inf,=,\pi)$-acceptance, since only 3-accepting $\omega$-grammars were considered in the literature. We showed that for some acceptance modes, the relative generative power of the $\omega$-grammars with respect to corresponding $\omega$-automata may be different from that of 3-accepting $\omega$-grammars. Third, the $\wHyp{CFG}$ with production repetition sets was studied to provide uniform relations and translation techniques over the entire hierarchy of $\omega$-grammars. However, the literature only considered $\wHypV{CFG}$, and the variable repetition sets are not applicable to $\wHyp{CSG}$ and $\wHyp{PSG}$ for example. Fourth, we tried to provide uniform proofs for various acceptance modes, since some of the proofs in the literature are based on the $\omega$-Kleene closure of language families (again, related to 3-acceptance), rather than uniform constructive proofs over various acceptance modes, e.g., the proof for $\epsilon$-production-free 3-accepting $\wHypV{CFG}$ (see Thm. 4.2.2 - Thm. 4.2.5 in \cite{CG77a}). Fifth, we considered one more acceptance mode beyond the five types of $i$-acceptance for $\omega$-automata, for the correspondence between $\omega$-grammars and $\omega$-automata.

Later, Engelfriest studied the six types of $(\sigma,\rho)$-accepting $\omega$-automata from the perspective of $X$-automata \cite{EH93}, but did not consider the grammar form. It is reasonable to establish the corresponding grammar forms for these $\omega$-automata. Therefore, we proposed $(\sigma,\rho,\pi)$-accepting $\omega$-grammars corresponding to $(\sigma,\rho)$-accepting $\omega$-automata, and established the relationship and translation techniques between $\omega$-grammars and $\omega$-automata.

\section{Conclusion}
\label{Sec:conc}
This paper married the related works on $\omega$-automata in the literature, and proposed the $(\sigma,\rho,\pi)$-accepting $\omega$-grammar. The relative generative power of $\omega$-grammars w.r.t. $\omega$-automata has been systematically studied, and compared in Table \ref{Tab:GenPow_wG_wA}.
\begin{table}[htb]
  {\centering
  \begin{tabular}{c|cccc}
    \hline
    ~                              & $\wHyp{RLG}$ & $\wHyp{CFG}$ & $\wHyp{CSG}$ & $\wHyp{PSG}$ \\
    \hline
    $\mathcal{L}_{\sigma,\rho,l}$  & $\mathcal{L}_{\sigma,\rho}(\wHyp{FSA})$ & $\mathcal{L}_{\sigma,\rho}(\wHyp{PDA})$ & $\subseteq \mathcal{L}_{\sigma,\rho}(\wHyp{PDA})^{(1)}$ & $\mathcal{L}_{\sigma,\rho}(\wHyp{PDA})$ \\
    $\mathcal{L}_{\sigma,\rho,nl}$ & $\mathcal{L}_{\sigma,\rho}(\wHyp{FSA})$ & $\subseteq \mathcal{L}_{\sigma,\rho}(\wHyp{PDA})^{(2)}$ & $\mathcal{L}_{\sigma,\rho}(\wHyp{TM})$ & $\mathcal{L}_{\sigma,\rho}(\wHyp{TM})$ \\
    \hline
  \end{tabular}
  }\\
  (1) $\mathcal{L}_{\sigma,\rho,l}(\wHyp{CSG}) = \mathcal{L}_{\sigma,\rho}(\wHyp{PDA})$, for $(\sigma,\rho) \not\in \{(\ran, \sqcap),(\ran, =)\}$.\\
  (2) $\mathcal{L}_{\sigma,\rho,nl}(\wHyp{CFG}) \subset \mathcal{L}_{\sigma,\rho}(\wHyp{PDA})$, for $(\sigma,\rho) \in \{ (\inf,\sqcap), (\inf,=)\}$.\\
      $\mathcal{L}_{\sigma,\rho,nl}(\wHyp{CFG}) = \mathcal{L}_{\sigma,\rho}(\wHyp{PDA})$, for $(\sigma,\rho) \in \{ (\ran,\subseteq), (\inf,\subseteq)\}$.
  \caption{Relative Generative Power of $\omega$-Grammars w.r.t. $\omega$-Automata}\label{Tab:GenPow_wG_wA}
\end{table}

One should particularly note:
\begin{enumerate}
  \item the generative power of $\wHyp{CFG}$ may be strictly weaker than or equal to that of $\wHyp{PDA}$.
  \item the generative power of leftmost derivations of $\wHyp{CSG}$ or $\wHyp{PSG}$ is not greater than that of $\wHyp{PDA}$.
  \item the generative power of $\wHyp{CSG}$ is equal to $\wHyp{TM}$. As a result, it is not necessary to define linear-bounded $\omega$-automata-like devices, since $\wHyp{CSG}$ does not have an independent level of generative power.
\end{enumerate}
We only relate $\omega$-grammars and $\omega$-automata in each acceptance mode separately, but actually many of these classes are equal. These equivalence relations can be easily inferred by using Theorem \ref{Thm:Gen_X_wAutomata}. This work is left to the interested reader.

The open problems lie on the $(\ran, \sqcap)$-acceptance and $(\ran, =)$-acceptance modes. Fortunately, the known results are enough for major applications, which concern mainly $(\inf, \rho)$-acceptances.

Thanks to the theorems and the translation techniques developed in the proofs in this paper, the closure property and the decision problems of the families of $\omega$-languages generated by $\omega$-grammars and those of the families of $\omega$-languages recognized by $\omega$-automata can be deduced from each other.

\section*{Acknowledgement}
This work was supported by the China Postdoctoral Science Foundation and the National Natural Science Foundation of China (60703033 and 61033002). The author wishes to thank the editor and the anonymous referees for their detailed comments and helpful suggestions.

\bibliographystyle{fundam}
\bibliography{fundam-chen}

%%%%%%%%%%%%%%%%%%%%%%%%%%%%%%%%%%%%%%%%%%%%%%%%%%%%%%%%%%%%%%%%%%%%%%

\end{document}